\documentclass[a4paper,fleqn,usenatbib]{mnras}

\usepackage{mathptmx}
\usepackage{txfonts}
\usepackage[pdf]{pstricks}
\usepackage[T1]{fontenc}
\usepackage{ae,aecompl}

\usepackage{ulem}
\usepackage{color} 
\usepackage{xcolor}

\usepackage{graphicx}	
\usepackage{amssymb}	

\newcommand{\Msun}{$M_{\odot}$ }
\newcommand{\fCBM}{$f_{\rm CBM}$ }
\newcommand{\fPDCZ}{$f_{\rm PDCZ}$ }

\newcommand{\fPDD}{$f_{\rm PDCZ}$ }
\newcommand{\fCE}{$f_{\rm CE}$ }
\newcommand{\fCHB}{$f_{\rm CHB}$ }
\newcommand{\fCHeB}{$f_{\rm CHeB}$ }

\newcommand{\Iso}[2]{$^{\rm #1}{\rm #2}$}
\newcommand{\ratioIso}[4]{\Iso{#1}{#2}/\Iso{#3}{#4}}

\title[Convective Boundary Mixing on the TP-AGB]{Impact of Convective Boundary Mixing on the TP-AGB}

\author[G. Wagstaff et al.]{
G. Wagstaff,$^{1,2}$\thanks{E-mail: wagstaff@mpa-garching.mpg.de}
M. M. Miller Bertolami,$^{3,4}$\thanks{E-mail: marcelo@mpa-garching.mpg.de}
A. Weiss,$^{1}$
\\
$^{1}$Max-Planck-Institut f\"{u}r Astrophysik, Karl-Schwarzschild-Str. 1, 85748 Garching, Germany.\\
$^{2}$ Exzellenzcluster ``Universe'', Technische Universit\"at M\"unchen,
Boltzmannstr. 2., D-85748 Garching, Germany.\\
$^{3}$ Instituto de Astrof\'isica de La Plata, UNLP-CONICET, Paseo del Bosque s/n, 1900 La Plata, Argentina.\\
$^{4}$ Facultad de Ciencias Astron\'omicas y Geof\'{\i}sicas, UNLP, La Plata, Argentina. \\
}

\date{Accepted 31.02.2021 Received 30.02.2020; in original form 29.02.2019}

\pubyear{2019}

\begin{document}
\label{firstpage}
\pagerange{\pageref{firstpage}--\pageref{lastpage}}
\maketitle

\begin{abstract}

The treatment of convective boundaries remains an important source of
uncertainty within stellar evolution, with drastic implications for
the thermally-pulsing stars on the Asymptotic Giant Branch
(AGB). Various sources are taken as motivation for the incorporation
of convective boundary mixing during this phase, from s-process
nucleosynthesis to hydrodynamical models. In spite of the considerable
evidence in favour of the existence of convective boundary mixing on
the pre-AGB evolution, this mixing is not universally included in
models of TP-AGB stars.  The aim of this investigation is to ascertain
the extent of convective boundary mixing, which is compatible with
observations when considering full evolutionary models. Additionally,
we investigate a theoretical argument that has been made that
momentum-driven overshooting at the base of the pulse-driven
convection zone should be negligible. We show that, while the argument
holds, it would similarly limit mixing from the base of the convective
envelope. On the other hand, estimations based on the picture of
turbulent entrainment suggest that mixing is possible at both
convective boundaries. We demonstrate that additional mixing at
convective boundaries during core-burning phases prior to the
thermally-pulsing Asymptotic Giant Branch has an impact on the later
evolution, changing the mass range at which the third dredge-up and
hot-bottom burning occur, and thus also the final surface
composition. In addition, an effort has been made to constrain the
efficiency of convective boundary mixing at the different convective
boundaries, using observational  constraints. Our study suggests
a strong tension between different  constraints that makes it
impossible to reproduce all observables simultaneously within the
framework of an exponentially decaying overshooting.  This result
  calls for a reassessment of both the models of convective boundary
  mixing and the observational constraints.

\end{abstract}





\section{Introduction}

The treatment of mixing at convective boundaries presents a challenge
which, despite the advent of 3D hydrodynamical simulations
\citep{Hurlburt1994, Freytag1996, Herwig2007, Baraffe2017}, has proven
to be a persistent source of uncertainties in 1D stellar evolution
models. This is a fact which can become especially relevant when
considering the evolution during the thermally pulsing Asymptotic
Giant Branch (TP-AGB), where boundaries at the edge of convective
zones are known to play an important part in governing many important
properties and observables of these stars
\citep{Herwig1997,Herwig2005}, such as third dredge-up (TDU) and the
Initial-Final Mass Relation (IFMR). Certainly, it has been seen in
hydrodynamic simulations \citep{Freytag1996, Herwig2007, Meakin2007,
  Mocak2009, Baraffe2017, Pratt2017} that the strict Schwarzschild
boundary, as implemented in 1D stellar evolution codes, simply does
not appear in the form of a composition discontinuity in a spherically
symmetric manner when it comes to multi-dimensional models.

Common ways of including additional mixing at convective boundaries
are either the extension of the fully mixed convective region by a
fraction of the local pressure scale height
\citep[e.g.][]{1992A&AS...96..269S} or the inclusion of an exponentially
decaying velocity profile \citep[e.g.][]{Herwig1997}. While it is well
established \citep{Maeder1991, Stothers1992, Schroder1997,
  Ekstrom2012} that some additional mixing is required from convective
cores on the upper main sequence, it is not always the case that
models produced for the TP-AGB (such as for yields or evolutionary
tracks) include significant, or any, convective boundary mixing (CBM)
from convective cores during the previous evolution
\citep{Cristallo2009,Karakas2007}. One aspect of this paper is to
focus on this pre-AGB treatment of CBM, to ascertain whether this is
having a significant effect on the models which are then
produced. This is of relevance given the output from these grids of
models are utilized by the wider astrophysics community. Additionally,
the application of the intensity of CBM calibrated from core hydrogen
burning stars to all convective boundaries can lead to results which
are in tension with observations when applied to all boundaries during
the TP-AGB evolution. This has been shown to be the case for the
stellar evolution codes used here
\citep{Weiss2009,Andrews2015,Bertolami2016}, and provides an
additional motivation for examining the influence of additional mixing
at the various convective boundaries. However, it is also not as
simple as preventing any CBM during the TP-AGB phase, as the existence
of carbon stars (C-stars) and the abundances of the PG1159 post-AGB 
stars provide strong evidence that CBM is indeed necessary. On the
theoretical side, \cite{Lattanzio2017} pointed out that traditional,
momentum-based overshooting cannot be the cause of additional CBM at
the base of the Pulse-Driven Convection Zone (PDCZ). However, as
presented by \cite{Meakin2007} convective boundary mixing in phases of
high thermal imbalance, such as the thermal pulses, might take the
form of turbulent entrainment. We explore both arguments in the
present work.

The aim of this paper is to test the impact of pre-AGB CBM on
  TP-AGB models, and to assess the calibration of TP-AGB CBM through
  the use of  the GARching STellar Evolution Code \citep[{\tt
    GARSTEC}]{Weiss2008} and the La Plata stellar evolution code
\citep[e.g.][{\tt LPCODE}]{Bertolami2016}, and thereby to improve
stellar models in this evolutionary phase. To begin with, \S \ref{sec:
  CBM} discusses the treatment of CBM in our stellar evolution codes
and presents the main convective boundaries in low-mass stellar
evolution. A discussion of the argument presented in
\cite{Lattanzio2017} together with an exploration of the possibility
of CBM by turbulent entrainment then follows in \S \ref{sec: Max
  OS}. \S \ref{sec:numex} and \ref{sec: TP-AGB mixing} then presents our numerical experiments and discusses the
implication for additional mixing at different convective boundaries
and the possibility of constraining CBM by means of various
observables. Finally, \S \ref{sec:remarks} ends the paper with some concluding
remarks.

\section{Convective Boundary Mixing}
\label{sec: CBM}

Traditionally referred to as overshooting, Convective Boundary Mixing
(CBM) has become a more commonly used term to account for the
uncertainty about the mechanism responsible for such mixing and to
acknowledge that it may not, in all cases, be momentum-based
overshooting. In {\tt GARSTEC} and {\tt LPCODE}, CBM is treated
  diffusively, according to the description of \cite{Freytag1996},
  where the diffusion constant at a distance $z$ from the convective
  boundary is given by
\begin{eqnarray}
\label{Eq: CBM}
 D(z) = D_0 \exp \left( \frac{-2z}{f_{\rm CBM}H_{\rm P}} \right)
\end{eqnarray}
where $f_{\rm CBM}$ is a free parameter that can be different at each
convective boundary and $D_0$ is the value of the diffusion constant
close to the convective boundary,  derived from the convective
  velocity. Within mixing-length theory $D_0$ is exactly zero by definition at the
formal convective border; therefore different codes define its value in
different ways:  The {\tt GARSTEC} version used for the present
  work, adopts $D_0$ as the value of $D$ at the convective grid point
  next to the formal convective boundary, while in {\tt LPCODE} $D_0$
  is defined as the mean value within $0.1 H_P$ of the formal
  convective boundary. In addition to $D_0$, a cutoff value needs to
  be implemented in stellar evolution codes. This cutoff value is
  taken as $D^{\rm cut off}=10^{-20}D_0$ ($D^{\rm cut
    off}=10^{-10}D_0$) in {\tt GARSTEC} ({\tt LPCODE}).

It is very difficult to determine how much additional mixing is likely
needed in stellar evolution models at different convective boundaries
and as such a variety of methods are used for current TP-AGB
evolutionary models. These range from nucleosynthesis arguments for
the extent of CBM necessary to reproduce the \Iso{13}{C} pocket below
the convective envelope to hydrodynamical simulations
 or AGB/post-AGB observables that need to be fitted by a given choice of the CBM intensity. These arguments are subsequently reviewed here.

\subsection{Core Hydrogen Burning}
There is considerable evidence for additional mixing for Core Hydrogen
Burning stars (CHB) on the upper main sequence, with a consensus that
it extends 
to roughly 20\% of the local pressure scale height
\citep{1992A&AS...96..269S, Herwig1997, Weiss2009,Bertolami2016,
  Ekstrom2012}. This is seen by the ability to reproduce the
observations of the main sequence stars in open clusters
\citep{Maeder1991, Stothers1992, Schroder1997} and eclipsing binaries
\citep{Claret2007, Stancliffe2015, Higl2017}.

\subsection{Core Helium Burning}

 CBM at the He-burning core can be physically motivated
  \citep{1971Ap&SS..10..340C,Castellani1985}. \cite{1971Ap&SS..10..340C}
  showed that any extension of the convective boundary beyond its
  formal value as given by the simple \citep[and naive,
    see][]{2014A&A...569A..63G} application of the Schwarzschild
  criterion is expected to increase the C abundance of the
  neighbouring layers. This in turn leads to an increase in
  their opacity, and consequently of $\nabla_{\rm rad}$, and thus to an even
  larger convective core.  The increase of the size of the
  convective core moves the convective boundary, and CBM, even further
  until $\nabla_{\rm rad}$ equals the local value of the adiabatic
  gradient $\nabla_{\rm ad}$. In fact, it can be argued that a correct
  application of the Schwarzschild criterion should always guarantee
  that neutral buoyancy (i.e. $\nabla_{\rm rad}=\nabla_{\rm ad}$) is
  attained at both sides of the convective border
  \citep{2014A&A...569A..63G,2017RSOS....470192S}.  In addition to
  this self-driving mechanism, the subsequent He-core burning later gives rise
  to the appearance of splittings in the formal convective core that
  can be modelled as a partially mixed region, where neutral buoyancy
  is attained \citep{Castellani1985}. The inclusion of some minor CBM
  in stellar evolution models allows the convective zones to stay
  connected, and although details in the final chemical profiles keep
  a record of the exact method adopted for computing mixing beyond the
  formal convective boundary, under moderate assumptions for CBM all
  algorithms lead to similar sizes in the final He-burning cores
  \citep{2015MNRAS.453.2290B,constantino2015,Constantino2017,2019A&A...630A.100D}.

Asteroseismic studies of Core Helium Burning (CHeB) stars support the
existence of the processes mentioned above \citep{2011A&A...530A...3C,
  constantino2015}, and might even help to select a preferred
algorithm \citep{2015MNRAS.453.2290B}. The evidence from star counts
\citep{Constantino2016} also seems to suggest that the naive
implementation of the Schwarzschild Criterion  in CHeB wrongly
estimates the size of He-burning cores \citep[as pointed out by][]{2014A&A...569A..63G}.

Numerical experiments with {\tt LPCODE} show \citep[see Fig. 2
  of][]{2019A&A...630A.100D} that, as long as $f_{\rm CHeB}$ is
included and kept within a moderate range ($0<f\lesssim 0.035$) the
evolution of all relevant quantities remains almost unchanged. Due to
the lack of very tight constraints and the fact that under moderate
assumptions for CBM in the He-burning core all algorithms lead to
similar sizes in the convective cores
\citep{2015MNRAS.453.2290B,constantino2015,Constantino2017,2019A&A...630A.100D},
it is common practice to take the same calibration as in the core
hydrogen burning phase and to apply it to the core helium burning as
well \citep[e.g.][]{Weiss2009,Bertolami2016, Jones2016,Ritter2017}.

\subsection{Convective Envelope \& Pulse-Driven Convective Zone}

The boundaries of these convective zones are best considered together,
given the inevitable connection between the two when considering the
TP-AGB. It is already known that stellar evolution codes tend to
require some form of additional mixing on the TP-AGB in order to
achieve sufficient dredge-up to reproduce statistics from carbon star
counts (e.g. from \citealt{Girardi2007}), at the necessary mass and
metallicity \citep{Herwig1997, Weiss2009,Bertolami2016}. However, it
has also been shown that the inclusion of the value for \fCBM as
calibrated by upper main sequence stars at all convective boundaries
\citep[e.g. in][]{Weiss2009} results in such efficient third dredge-up
that core growth is suppressed during the TP-AGB, leading to an IFMR
that is in tension with observations, the predicted final mass
  being too low by $\sim 0.1...0.2 M_\odot$ for initial masses in the
  range $\sim 2...4 M_\odot$ 
\citep{Salaris2009,2018ApJ...866...21C,2018ApJ...860L..17E}.

 CBM below the convective envelope (CE) plays two different
  roles. On the one hand even the inclusion of an inefficient CBM at
  the bottom of the CE is known to enhance TDU during the TP-AGB
  \citep{Herwig2000}. On the other hand, CBM below the CE is required
  to enrich sufficiently the partially mixed regions in H and C to
  ensure the formation of a $^{13}$C pocket. The radiative burning of this
  $^{13}$C pocket is the main source of neutrons for the formation of
  heavy elements through slow neutron captures (s-process
  nucleosynthesis). Attempts to calibrate the value of
  \fCE by different means have lead to contradictory results. While
  some authors have found that very intense CBM might even be required
   to match the O-rich to C-rich transition luminosity of AGB stars in Magellanic Cloud clusters  \citep{2012ApJ...746...20K}, others
  concluded that such intense CBM at the bottom of the CE cannot be
  reconciled with the ratios of C- and M-stars in the Magellanic
  Clouds' Globular Clusters or the IFMR \citep{Bertolami2016}. In
  fact, several recent works do not include any CBM at all at the
  bottom of the CE for the computation of their model grids
  (i.e. \fCE=0, e.g. \citealt{2014MNRAS.445..347K, Bertolami2016}).  
This is in stark contradiction to what is necessary to reproduce
s-process abundances. From a calibration of the $^{13}$C pocket,
\cite{Cristallo2009} found an intensity of CBM at the bottom of the
convective envelope (CE) that corresponds to a value of $f_{\rm
  CE}\sim 0.2$. Similarly, from the partially mixed mass required at
the bottom of the convective envelope given by \cite{Lugaro2003} and
\cite{Herwig2003}, \cite{Ritter2017} derived a value of \fCE=0.126.
Note that in all these papers the diffusive approach for CBM was used.

Regarding CBM at the PDCZ, a combined 2D/3D study by \cite{Herwig2007}
tried to focus specifically on the TP-AGB boundaries. The result of
which suggested that the boundary at the base of the PDCZ could be
approximated by two exponential decays, the first beginning inside the
convective region with a value of $f_{\rm PDCZ}=0.01$, and a second
outside the convective zone with a value of $f_{\rm PDCZ}=0.14$. The top
boundary was more simply modelled by a single decay, with a value of
\fCBM=0.1. However, in later works \cite{Ritter2017} suggest that a
value of \fPDCZ=0.008 is motivated by the same simulations of
\cite{Herwig2007}.

At least some additional mixing appears to be required from the
observations of post-AGB stars (specifically those referred to as
PG~1159 stars) which are believed to  display at the photosphere
the final intershell abundances of TP-AGB stars \citep{Herwig1999}. A
constraint of \fPDCZ=0.01-0.03 was suggested from a preliminary
investigation carried out by \cite{Herwig2000}. These results were
extrapolated from a few thermal pulses of a single evolutionary model,
but later explorations \citep{MB15, Bertolami2016} suggest a lower
value, \fPDCZ$\sim 0.005$--$0.0075$.

\section{Characterising convective boundaries on the TP-AGB}
\label{sec: Max OS}

\subsection{A buoyancy argument and momentum-driven overshoot}
Besides the observational and hydrodynamical constraints, some
physical arguments can be constructed to understand the situation of
CBM during the TP-AGB.

\begin{table*}
\begin{tabular}{cccccc}\hline
Convective Border & $v_{\rm MLT}$  & ${\rm Pe}$      &  $d_{\rm pen}$    & $d_{\rm ent}/H_P^0$ & $\tau^{\rm 1HP}$ \\
                  & (cm/s)         &           &  (cm, $H_P$)       &                      & (yr)     \\\hline
 \multicolumn{6}{c}{$6^{\rm th}$TP: $\log L_{\rm He}^{\rm max}=6.85$, $\tau^{\rm peak}=2.5$yr} \\
 Bottom PDCZ & $2.25\times 10^5$  & $10^7$ & 3950, $2.2 \times 10^{-5}$&  0.16     & 25          \\
 Bottom CE (Max. $L_{\rm He}$) & $8\times 10^4$  & $500...3000$ & $2.5\times 10^7$, $3.3 \times 10^{-3}$&  0.37     & 68  \\
 Bottom CE (Max. Depth CE)& $4.3\times 10^5$  & $1000...10000$ & $1.6\times 10^6$, $5.9 \times 10^{-4}$&  -     & 4.1  \\\hline
  \multicolumn{6}{c}{$13^{\rm th}$TP: $\log L_{\rm He}^{\rm max}=8.23$, $\tau^{\rm peak}=0.127$yr} \\
 Bottom PDCZ & $6\times 10^5$  & $10^7...10^8$ & 3500, $2.2 \times 10^{-5}$&  0.11     & 1.4          \\
 Bottom CE (Max. $L_{\rm He}$)& $9.6\times 10^4$  & $700...2000$ & $2\times 10^7$, $3.3 \times 10^{-3}$&  0.17     & 32  \\
 Bottom CE (Max. Depth CE)& $5.7\times 10^5$  & $350...4000$ & $1.3\times 10^6$, $3.5 \times 10^{-4}$&  -     & 1.9  \\\hline 
\end{tabular}
\caption{Characteristic values of the turbulent velocity $\nu_{\rm
    MLT}$, P\'eclet Number ${\rm Pe}$, penetration distance $d_{\rm
    pen}$, distance entrained during the duration of the flash $d_{\rm
    ent}$, and timescale required to entrain a distance equal to the
  local pressure scale height $\tau^{\rm 1HP}$, for different
  convective boundaries and during the 6th and 13th thermal pulses of
  an initially $M_i=3 M_{\odot}$ sequence ($Z_{\rm ZAMS}=0.01$)
  computed with {\tt LPCODE} \citep{Bertolami2016}; see text for
  additional explanations. }
\label{tab:characteristic_numbers}
\end{table*}

Thermal pulses on the AGB develop when the He-burning shell becomes
geometrically thin and the gravothermal specific heat becomes
positive, leading to a thermal runaway (see section 34.2 in
\cite{Kippenhahn2012} for a detailed discussion of the process). As a
consequence, during a thermal pulse temperature rises, leading to an
increase by several orders of magnitude of the He-burning luminosity,
which reaches typical values of $L_{\rm He}\gtrsim 10^7 L_\odot$. This
leads to the development of a convective zone (the PDCZ) that extends
outwards from the temperature peak. In addition of the creation of the
PDCZ itself, the thermal pulse also creates a significant temperature
inversion at the lower convective boundary of the PDCZ.
An argument has been made \citep{Lattanzio2017} that this temperature
inversion would act so strongly against any convective eddy which may
emerge from the bottom of the convective zone, that it would be
impossible for any momentum-driven overshooting to cover any
appreciable distance. The basic physical argument is outlined in
\cite{Lattanzio2017}, and it focuses on the buoyancy acting against
an convective element which travels beyond the formal convective
boundary. By considering a convective element of density $\rho_e(r)$
at the base of the convective zone, moving downwards adiabatically,
they compute the deceleration of the element until it stops. Then the
distance $d_{\rm pen}=|r_1-r_0|$ traveled beyond the formal convective
boundary (at $r_0$) is given by
\begin{eqnarray}
\label{eq: PDCZ OS}
\frac{1}{2} v_0^2 = \int_{r_0}^{r_1} g(r) \left[ \frac{\rho(r) - \rho_e(r)}{\rho_e(r)}\right] dr
\
\end{eqnarray}
where $v_0$ is a ``typical turbulent velocity'' near the convective
boundary. It should be mentioned that the adiabatic approximation is
very accurate for convective elements in deep convective zones such as
the PDCZ. Also at the bottom of the convective envelope of red giants
the adiabatic approximation is good, as discussed by
\cite{2015A&A...580A..61V}. This is reflected by the large Pecl\'ec
numbers (${\rm Pe}$) of convective elements in both the PDCZ and the
CE --see Tab. \ref{tab:characteristic_numbers}. Above argument has
been applied to models calculated with {\tt GARSTEC} and {\tt LPCODE}
to investigate its relevance. At variance with what is done by
\cite{Lattanzio2017}, we estimate the value of $v_0$ near the edge of
the convective zone (we take $v_0$ as the mean value within $0.1 H_P$
of the convective boundary) instead as the maximum value within the
convective zone. This leads in our case to values of $d_{\rm pen}$
lower by a factor of a few in comparison with those estimated by
\cite{Lattanzio2017}.

 Table~\ref{tab:characteristic_numbers} shows characteristic
  numbers during the 6th and 13th thermal pulses of an initially
  $M_i=3 M_{\odot}$ sequence ($Z_{\rm ZAMS}=0.01$) computed with {\tt
    LPCODE}. Models are taken from \cite{Bertolami2016} but
  with CBM inhibited for this experiment.  As shown in Table
\ref{tab:characteristic_numbers},  according to the criterion
  presented by \cite{Lattanzio2017} convective elements should 
    indeed barely
  penetrate into the neighboring stable regions. The estimated
penetrated distance is of the order of $10^{-5}\times H_P$ for the
lower boundary of the PDCZ and somewhat larger,  but still
  very small at the bottom of the CE
(around $10^{-4}\times H_P$ at the point of maximum penetration of the
CE).

\subsection{Stiffness of the convective boundary: $Ri_{\rm B}$}
\label{sec: RiB}
The previous dynamical argument suggests that no momentum-driven
overshoot at the bottom of any convective zone should be expected
during the TP-AGB. However, as noted by \cite{2015A&A...580A..61V} in
quick phases of high thermal imbalance the proper regime of CBM might
correspond to that of turbulent entrainment as defined by
\cite{Meakin2007}. In the case of turbulent entrainment, turbulence
diffuses into the stable regions leading to a progressive advance of
the whole border of the turbulent region. Consequently, while the argument discussed above might suggest the
absence of overshooting at TP-AGB convective boundaries, it might be
far from the actual physical process driving convective boundary
mixing in these stars. According to \cite{Meakin2007} the speed $u_E$
at which the turbulent border advances into the neighboring stable
layers is given by
\begin{equation}
u_E=\sigma\, A\, {\rm Ri}_{\rm B}^{-n}
\end{equation}
where ${\rm Ri}_{\rm B}$ is the bulk Richardson number and $\sigma$
the rms of velocities at the turbulent border. A fit to numerical
simulations of oxygen-shell burning and core-hydrogen burning by
\cite{Meakin2007} suggests that $\log A=0.027\pm 0.38$ and $n=1.05\pm
0.21$, in agreement with theoretical arguments and other numerical
experiments. For all practical purposes this means that $u_E\simeq
\sigma/ {\rm Ri}_{\rm B}$. The bulk Richardson number is a measure of
the stiffness of the boundary region, defined as
\begin{equation}
{\rm Ri}_{\rm B}=\frac{\Delta b\, L}{\sigma^2},
\end{equation}
where $L$ is a length scale of the turbulent motions and $\Delta b$ is
the buoyancy jump across the transition. Here the relative buoyancy
$b(r)$ is defined by
\begin{equation}
b(r)=\int_{r_0}^r N^2 dr,
\end{equation}
where $N$ is the local angular buoyancy frequency.  Unfortunately, in
stellar evolution models the computation of $\Delta b$ cannot be
performed as the width of transition region between stable and
turbulent layers is undefined.

However, we can estimate a typical timescale that it
would take the boundary to reach a given radius $r$ beyond the
convective boundary as
\begin{equation}
\tau(r)=\frac{|r-r_0|}{u_E(r)}\simeq \frac{|r-r_0| b(r) L}{\sigma^3}
\label{eq:def_taur}
\end{equation}
where we have taken $r_0$ as the location of the formal convective
boundary in the stellar evolution model so that $b(r)=\Delta b$. By
estimating the length scale of turbulence as the mixing length at the
convective border ($L=\alpha_{\rm MLT} H_P^0$), and $\sigma$ as the MLT
velocity near the convective boundary (again we take $v_0$ as the mean
value within $0.1 H_P^0$ of the convective boundary), we can compute a
value for $\tau(r)$. In this way we can now estimate the entrainment
distance ($d_{\rm ent}$) of the turbulent border by equating the
$\tau(r)$ with a typical duration of the thermal pulse $\tau_{\rm
  peak}$. In what follows we define $\tau_{\rm peak}$ as the time
spent by the He-burning luminosity within one order of magnitude of
its maximum value\footnote{Note that as soon as the He-burning
  luminosity drops significantly, also the convective velocities drop
  and the formal convective boundary recedes.}.  In
Table~\ref{tab:characteristic_numbers} we show the entrainment
distances during two thermal pulses of an initially $M_i=3 M_\odot$ star
($Z=0.01$, taken from \cite{Bertolami2016}, but with CBM inhibited for
this experiment). At variance with the previous estimation, this
consideration suggests that the bottom of the
PDCZ is able to penetrate a distance of $d_{\rm ent}\gtrsim 0.1 H_P^0$
into the stable layers. If this is the case, then convective boundary
mixing at the bottom of the PDCZ is not only possible but also
significant, and should be included in stellar evolution models.

Additionally, we show in Table~\ref{tab:characteristic_numbers} the
time estimate ($\tau^{\rm 1HP}$) for the entrainment to penetrate a
distance of $d_{\rm ent}=1 H_P^0$ both at the bottom of the PDCZ and
the CE. While in the case of the PDCZ this timescale is about 10 times
the duration of the peak of He-burning, in the case of the bottom of
the CE the timescale is shorter than the duration of the CE being at maximum
depth after the thermal pulse (which is of $\sim 120$ yr and $\sim
320$ yr for the $3^{\rm rd}$ and $13^{\rm th}$ thermal pulses, respectively).  This
can be taken as a hint that convective boundary mixing at the bottom
of the CE at the time of maximum depth of the convective envelope
(which is responsible for the later formation of the $^{13}$C pocket
in AGB stars) is expected to be large.  From these considerations we
   conclude that CBM at the bottom of the CE is much
  more efficient than at the bottom of the  PDCZ. However a  word of
  caution is in order regarding the limitations of these estimations: 
While convection close to the border of the PDCZ has 
P\'eclet numbers\footnote{ Throughout this section we use the
  effective value of the P\'eclet number proposed by
  \cite{2015A&A...580A..61V} for stellar evolution models within the
  MLT theory of convection, which is ${\rm Pe}=3 D/\chi$ where $D$ and
  $\chi$ stand for the coefficients for chemical mixing and thermal
  diffusion respectively. } of the order of ${\rm Pe}\gtrsim 10^{7}$
 and the  entrainment law of \cite{Meakin2007} can thus be
  expected to be valid, 
values at the bottom of the convective envelope are typically more
than 4 orders of magnitude smaller (see
Table~\ref{tab:characteristic_numbers}). As a consequence, if the
velocity drops 
exponentially outside the formal convective border, it is expected that
 CBM  beyond the formal convective boundary will stop to be
adiabatic already close to the boundary, and the entrainment picture
of \cite{Meakin2007} will stop to be valid.

\section{Numerical Experiments}
\label{sec:numex}

Calculations in Sections~4 and 5 were performed mostly using {\tt
  GARSTEC} \citep{Weiss2008, Weiss2009, Wagstaff2018} and, when
explicitly noted, also {\tt LPCODE} \citep{Bertolami2016}.  Models
presented in Section~6 were computed with {\tt LPCODE}.  The physics
and numerics adopted in the present work are consistent with those
described in the above papers and the details can be found there.  The
exception to this is the main focus of study of this paper, i.e.\ the
efficiency of convective boundary mixing at different stages of the
evolution. For the sake of completeness we summarize some of the most
important ingredients adopted for the present work.  Both {\tt
  GARSTEC} and {\tt LPCODE} sequences where computed with an Eddington
gray atmosphere throughout the evolution. Both codes adopt
high-temperature radiative opacities from the OPAL project
\citep{Iglesias1996}, and low-temperature molecular opacities from
Wichita State Alexander \& Ferguson molecular opacity tables
\citep{Ferguson2005, Weiss2009} for solar-like as well as for C-rich
mixtures. Conductive opacities are included in this version of {\tt
  GARSTEC} \citep{Weiss2009} following \cite{1983ApJ...273..774I}
while {\tt LPCODE} adopts the conductive opacities of
\cite{2007ApJ...661.1094C}.  The neutrino emission by plasma processes
from the degenerate core of red giants is adopted in both codes
according to \cite{1994ApJ...425..222H}.  For the equation of state
(EOS) {\tt GARSTEC} sequences were computed with the Irwin's EOS
\citep{Cassisi2003} in the form of pre-calculated tables, while {\tt
  LPCODE} sequences adopt the updated {\tt EOS\_2005} version of the
OPAL EOS \citep{Rogers1996}. When outside the parameter range of
modern tabular EOS both codes rely on a Saha-type EOS. Regarding the
treatment of convective energy transport both codes rely on the mixing
length theory of convection (MLT) as described in
\cite{Kippenhahn2012}. The free parameter $\alpha_{\rm MLT}$ was taken
as $1.75$ and $1.822$ in {\tt GARSTEC} and {\tt LPCODE},
respectively. Mass loss has been included in similar fashion as in
previous works. {\tt GARSTEC} sequences were computed with a mixture
of \cite{Reimers1975}, \cite{Wachter2002}, and \cite{vanLoon2005}
prescriptions as described in \cite{Weiss2009}, while {\tt LPCODE}
sequences rely on a mixture of \cite{2005ApJ...630L..73S},
\cite{1998MNRAS.293...18G}, and \cite{2009A&A...506.1277G}, as
described in \cite{Bertolami2016}, where we have explored current
constraints on CBM during core hydrogen burning, core helium burning,
mixing from the lower boundary of a convective envelope, and from the
convective boundaries of pulse-driven convection zones.

  {\tt GARSTEC} models presented in Sections~4 and 5 were calculated for
  two compositions, taking Z=0.02 and Y=0.28 for the solar case and
  Z=0.008 and Y=0.25 for an LMC comparison. Both set of models take
  the abundances of \cite{Grevesse1993}.  TP-AGB models were computed
  until the computations failed to converge due to numerical
  instabilities just before the end of the TP-AGB
  \citep{Weiss2009,2012A&A...542A...1L}.   The values
    for the efficiency of CBM  in the following sections and experiments (namely, \fCHB, \fCHeB, \fCE, and
    $f_{\rm PDCZ}$) have been chosen to encompass the values derived by
    previous authors and discussed in Section~\ref{sec: CBM}.

\subsection{Convective Cores}
\label{sec: CB EvCalc Core}

The size of convective cores present in the pre-AGB stages of
evolution, during core hydrogen (CHB) and core helium burning (CHeB),
determine the mass of the H- and He-free cores at the beginning of the
TP-AGB. As a consequence, CBM at these convective cores can be
expected to play a role during the TP-AGB.  Exploring the impact of
CBM during CHeB and CHB stages on the TP-AGB evolution is the purpose
of this section. Here we compute models with {\tt GARSTEC} for
different initial masses that include, or exclude, CBM during the CHB
($f_{\rm CHB}=0.0174$) and/or the CHeB ($f_{\rm CHeB}=0.0174$) and
then follow the subsequent TP-AGB evolution. During the TP-AGB phase,
CBM was included only at the PDCZ ($f_{\rm PDCZ}=0.0075$, $f_{\rm
  CE}=0$), with the exception of an additional comparison sequence
that was computed without CBM at all convective boundaries. These
sequences are then labeled 'None', 'CHB-PDCZ', 'CHB-CHeB-PDCZ' and
'PDCZ' according to the convective zones at which CBM was
included. Note that the sequences label-led 'CHB-CHeB-PDCZ' have the
same choice of $f_{\rm CBM}$-parameters as the model grids computed by
\citet{Bertolami2016}.

\subsubsection{At the Onset of the TP-AGB}
\begin{figure}
  \centering
  \includegraphics[width=\columnwidth]{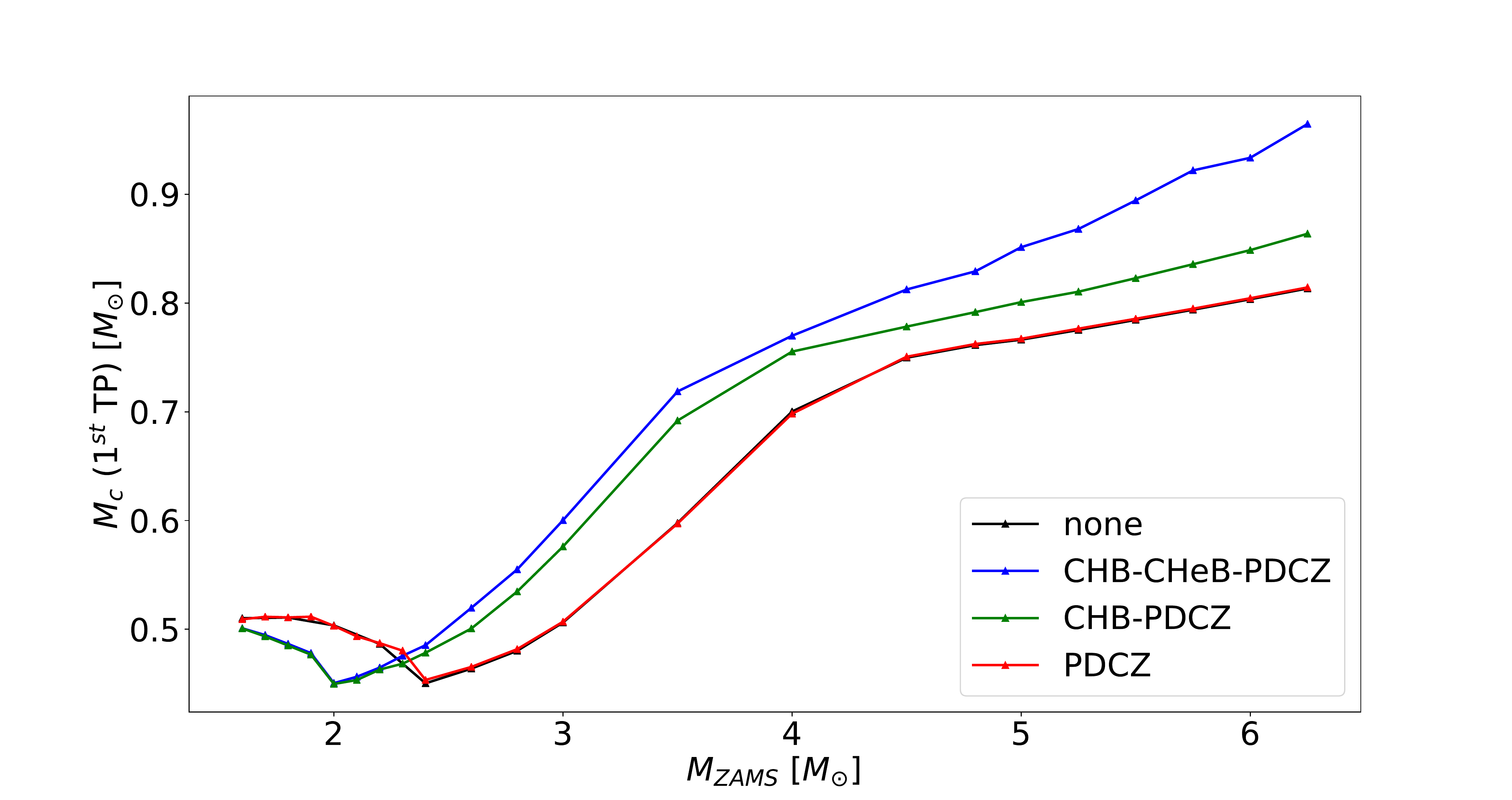}
  \caption[Core mass at the first thermal pulse for CBM]{The core mass
    at the first thermal pulse as a function of initial mass for
    sequences computed with CBM included at different convective
    boundaries.  All sequences correspond to initial metallicities
      of $Z=0.02$. Note the shift in $\sim 0.4 M_\odot$ in the
    progenitor mass when convective boundary mixing is included in the
    convective cores. } 
  \label{Fig: TP1M}
\end{figure}

 Fig.~\ref{Fig: TP1M} shows the mass of the H-free core
  ($M_c^{\rm TP1}$) at the moment of the first thermal pulse (i.e.\ the beginning
  of the TP-AGB phase) for stellar evolution sequences computed under
  different assumptions. Note that models with CBM only at the PDCZ
  boundaries and models with no CBM at any convective boundary are
  identical before the TP-AGB. The mass of the H-free core (HFC)
  provides a very clear indication of how CBM before the TP-AGB
  influences the behaviour on the TP-AGB.

The first thing that becomes apparent between models with and without
CBM during the core-burning stages is the shift in the minima of
$M_c^{\rm TP1}$ as a function of the initial mass $M_{\rm ZAMS}$
(Fig. \ref{Fig: TP1M}). This minima corresponds to the transition
initial mass below (above) which helium burning commences in a
degenerate (non-degenerate) core. This transition mass shifts by about
$\sim 0.4M_\odot$ when CBM is included during the main sequence
(CHB). Note that this transition mass is in no way affected by the
inclusion or not of CBM during the CHeB phase (see Fig.~\ref{Fig: TP1M} ). 
  This is because it is the size of the mixed core during the CHB
   that defines the size of the He-core after the main sequence, and
   whether such core is massive enough to ignite He in a
   non-degenerate way. CBM at the CHeB becomes only important once
   stable He-core burning begins.  The impact of CBM during the CHeB
   phase can be appreciated  for higher initial mass models, for
 which the inclusion of CHeB begins to influence $M_{\rm c}^{\rm TP1}$
 by increasing the core mass. This effect becomes even stronger with
 increasing initial mass.

The overall effect of including CBM for convective cores is that
models mimic the behaviour of models with larger initial masses
($\Delta M_i \sim 0.4 M_\odot$), but without CBM for convective
cores. This results in generally less massive cores for the low-mass
stars ($M_i\lesssim 2.5 M_\odot$), but more massive cores for higher
initial mass.  This will have a clear impact on the initial
  threshold masses $M_{\rm ZAMS}$ at which a star is expected to begin
  third dredge-up (TDU) or hot bottom burning (HBB).  Proper pre-AGB
  evolution is therefore important when AGB yields or AGB properties
  need to be linked to the initial progenitor masses, or the age of
  the host stellar population.

\subsubsection{Third Dredge-Up}

\begin{figure}
 \centering
  \includegraphics[width=\columnwidth]{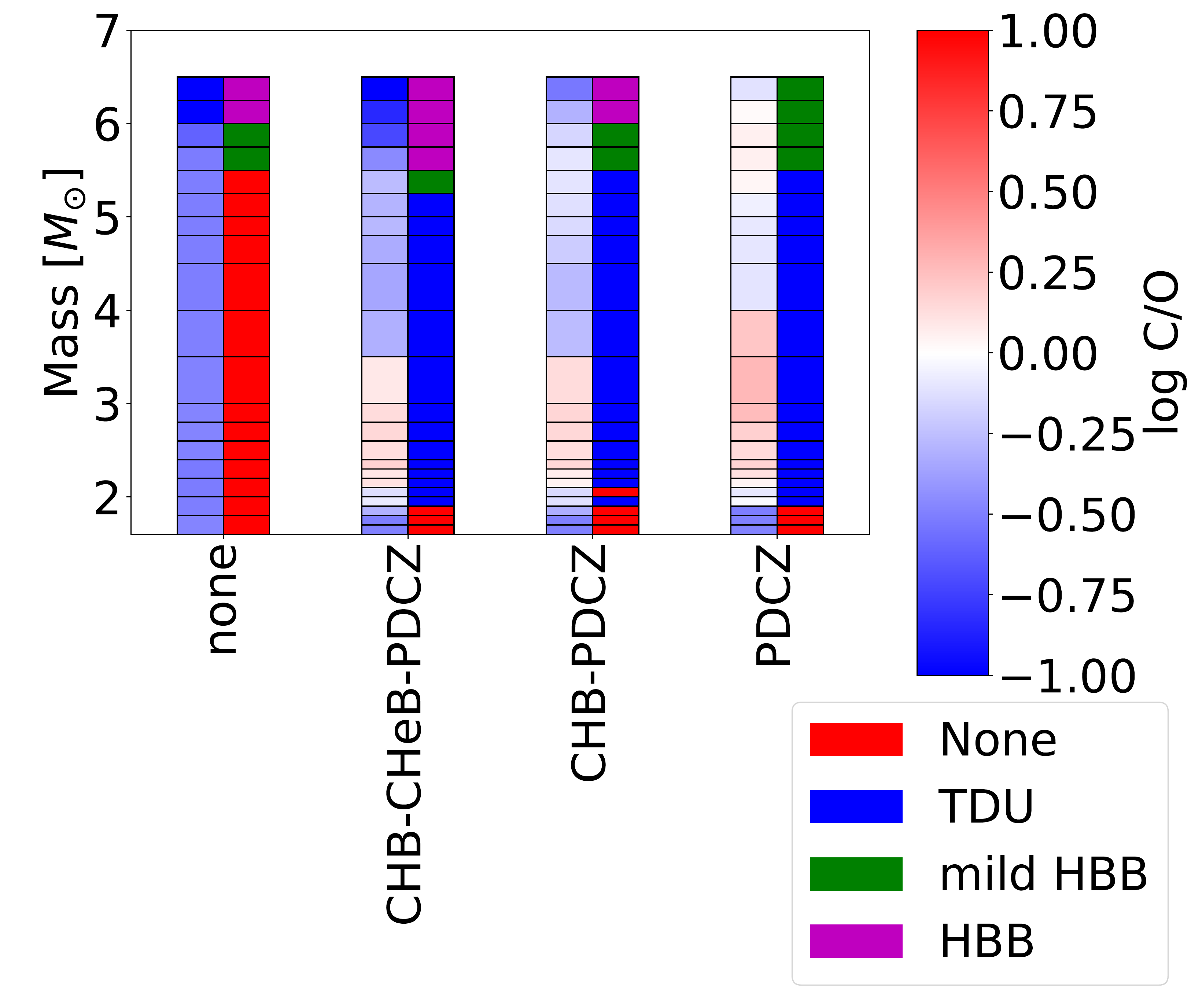}
  \caption{C/O values and classifications of TDU and hot-bottom
    burning (HBB) for  sequences that include CBM at different
    convective boundaries.  All sequences correspond to initial
      metallicities of $Z=0.02$. Each set of models is represented
    by two columns, the left of which shows the final surface C/O
    value (matched to the colour bar at the side) while the right hand
    column shows the model classification;  the initial mass is
      given along the vertical axis. The
    classification is defined as follows; Red -- no third dredge-up,
    blue -- third dredge-up, green -- mild hot-bottom burning, magenta
    -- hot-bottom burning. Initial C and O abundances correspond to $\log C/O=-0.32$ by number fraction.}
  \label{Fig: finalCO TPclass}
\end{figure}

Still focusing on the models including CBM during convective core
phases, the surface C/O values at the end of our computations can be
seen in Fig.~\ref{Fig: finalCO TPclass} as a function of initial mass,
for the different combinations of boundaries where additional mixing
has been applied. Alongside the final C/O value appears the
classification of whether the models undergo third dredge-up,
hot-bottom burning or what is here classified as mild hot-bottom
burning. This last category arises for models where a reduction in the
carbon isotopic ratio, \ratioIso{12}{C}{13}{C}, is observed although
no significant reduction in either the C/O or C/N value occurs.

In the set of models where no additional mixing is applied, there is a
lack of dredge-up in all models, although hot-bottom burning still
occurs for the higher mass models. In contrast, the models with mixing
only applied at the PDCZ experience a significant increase in carbon
due to dredge-up. The comparison between all the sets of sequences
that include CBM at the PDCZ clearly shows that core size has a
significant influence on hot-bottom burning, as  any difference
arises from the initial core size as was shown in Fig.~\ref{Fig:
  TP1M}.  This can be seen by the fact that the models with the
largest cores at higher masses, CHB-CHeB-PDCZ, experience hot-bottom
burning at the lowest initial masses and to the greatest degree, while the
CHB-PDCZ and PDCZ models undergo progressively less hot-bottom
burning. That the set of models with no additional mixing experience
hot-bottom burning to a larger extent than the PDCZ models, despite
beginning with the same core size, seems to be because mixing from the
PDCZ boundaries inhibits hot-bottom burning, due to the increase in
the helium luminosities generated.

As for third dredge-up, it can again be seen to alter its behavior as
a result of the treatment of convective core boundaries. The
similarity between the CHB-CHeB-PDCZ models and the CHB-PDCZ models
for initial masses which have degenerate cores at the onset of helium
burning, begins to diverge above this; the more efficient dredge-up
for the latter indicates that the lower core mass results in a final model
which is more carbon-rich. This pattern continues when considering the
PDCZ-only models, which have a lower initial core mass at higher
masses (and more efficient dredge-up), but larger cores at lower initial masses
(and less efficient dredge-up). Indeed, combined with the lack of
hot-bottom burning in the more massive PDCZ-only set of models, it can
even be observed that a 5.5\Msun model becomes carbon-rich.

Overall the effect of including additional CBM during core burning
phases can play a significant role in changing the outcome of TP-AGB
calculations,  and given the evidence in favour of the existence
  of CBM during the core burning stages this effect needs to be
  addressed. This evidence is also reinforced by the recent hints of the existence of HBB in stars with initial masses as low as $M_{\rm ZAMS}\sim 3
  M_\odot$\citep{2018MNRAS.473..241H,2018ApJ...853...50F,2019arXiv190908007D}.

  The inclusion of CBM in pre-AGB stages is particularly important
  when trying to link the C and N abundances measured in PNe with the
  initial mass of the progenitor star
  \citep[e.g.][]{2018MNRAS.473..241H,2018ApJ...861L...8M,2019arXiv190908007D}. In
particular, the adoption of nucleosynthesis computations based on
stellar models that do not include convective boundary mixing on the
upper main sequence
(e.g. \citealt{2015ApJS..219...40C,2016ApJ...825...26K}), might lead
to a systematic  shift in the nucleosynthesis predictions as a
  function of the initial mass by $\sim 0.4M_\odot$, see
Fig.~\ref{Fig: TP1M},  with additional consequences for deriving
  timescales of galactic chemical evolution.

\subsection{Impact of CBM during Second Dredge-up}
\label{sec: 2du CBM}

\begin{figure*}
 \centering
  \includegraphics[width=18cm]{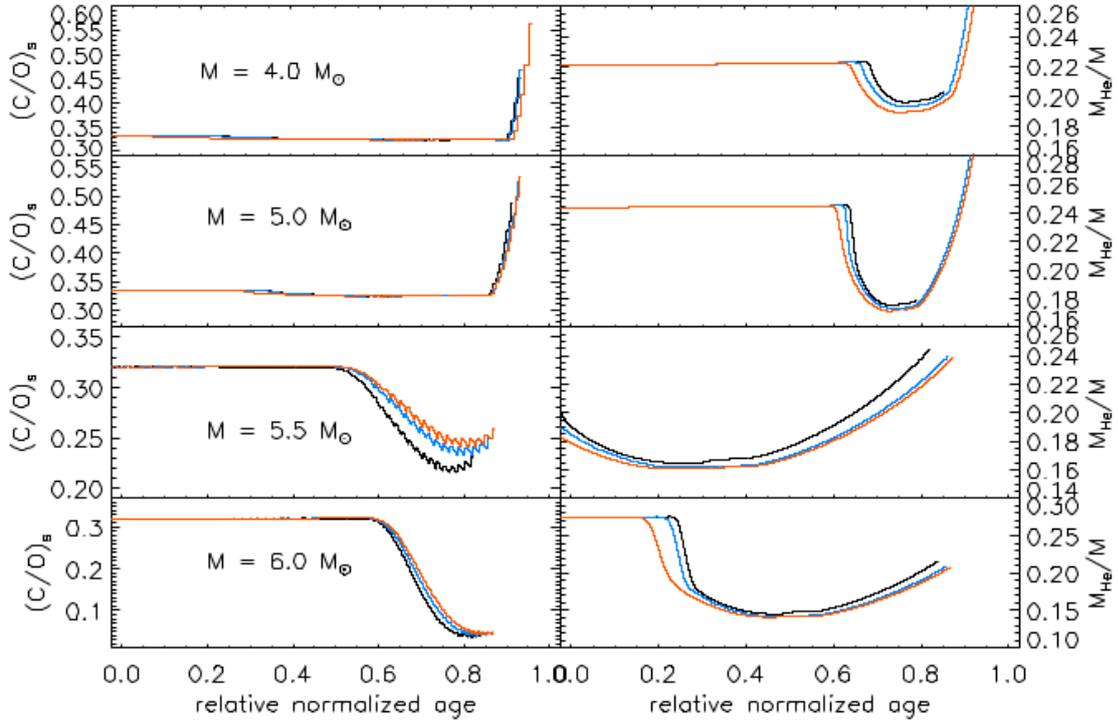}
  \caption{ Impact of CBM during 2DUP on the subsequent evolution
      for evolutionary sequences of $M_{\rm ZAMS}=4,\ 5,\ 5.5$, and $6
      M_\odot$. The sequences shown here have been computed in all
      cases with $f_{\rm CHB}=0.0174$,$f_{\rm CHeB}=0.0174$,$f_{\rm
        PDCZ}=0.0075$,$f_{\rm CE}=0$, during the main sequence,
      core-He burning and TP-AGB phases, only altering the value of
      $f_{\rm CE}$ during the 2DUP.  Left panels show the change in
      the photospheric C/O ratio when 2DUP is computed with strong
      (orange, $f_{\rm CE}=0.126$), moderate (blue,$f_{\rm
        CE}=0.0174$), or no CBM ($f=0$). Right panels show the
      evolution of the H-free core for the same evolutionary sequences
      (in mass coordinate). The relative normalized age plotted in the
      x-axis is defined in such a way that it covers the relevant
      period between the end of core helium burning and the end of the
      computation, in order to allow a direct comparison for all four
      mass values. For each mass value this ``relative normalized
      age'' is identical for all three 
      cases of $f_{\rm CE}$.  All sequences correspond to initial
      metallicities of $Z=0.02$.} 
  \label{Fig:2dup}
\end{figure*}

 In this section we analyze the impact of second dredge-up (2DUP)
  on the mass of the H-free core during the subsequent TP-AGB
  evolution, and on the final C/O ratio. In particular we try to
  assess the impact of 2DUP on the threshold for
  HBB. Fig.~\ref{Fig:2dup} shows the evolution of several sequences in
  which the 2DUP was computed under the assumption of different values
  of the CBM parameter $f_{\rm CE}$. Note that for the sake of these
  experiments, CBM during the main sequence, core-He burning, and
  TP-AGB phases was kept constant with values of $f_{\rm CHB}=0.0174$,
  $f_{\rm CHeB}=0.0174$, $f_{\rm PDCZ}=0.0075$, and $f_{\rm CE}=0$,
  respectively. Thus our reference sequence with no CBM at the bottom
  of the CE corresponds to those sequences labeled as CHB-CHeB-PDCZ in
  Fig.~\ref{Fig: finalCO TPclass}. The right panels in
  Fig.~\ref{Fig:2dup} show the evolution of the H-free core (and
  consequently the location of the bottom of the H-rich convective
  envelope) during the 2DUP and the TP-AGB. The incorporation of CBM
  during 2DUP produces a decrease of the H-free core of a few percent
  in $M_{\rm HFC}$ (up to $\sim 5$\%) at the point of maximum depth,
  i.e. at the beginning of the TP-AGB. While this difference is not
  negligible it is smaller than those produced by CBM during the main
  sequence. This becomes clear when comparing Fig.~\ref{Fig: TP1M}
  with Fig.  \ref{Fig:2dup}:  While CBM during the main sequence
    leads to differences of almost 20\% in the size of the H-free
    core, CBM at the CE during the 2DUP leads to initially only a
    change of a few percent, and even after the TP-AGB the impact of
    CBM during the 2DUP leads to differences of not more than 10\% in
    the H-free core size. Note that in these experiments CBM is
    included at the bottom of the CE only during 2DUP, but not during
    the TP-AGB.  Changes introduced by CBM during the 2DUP are also
  small regarding the impact of CBM during the 2DUP on the C/O ratio,
  and in particular of the development of HBB. As shown in the left
  panels of Fig.~\ref{Fig:2dup}, HBB is active for stars with
  $M_i\gtrsim 5.5 M_\odot$, noticeable by the continuous decrease in
  the C/O ratio, regardless of the assumption about CBM during the
  2DUP. This is in contrast to the dependence of HBB to the inclusion
  of CBM during the CHB, CHeB and PDCZ shown in Fig.~\ref{Fig: finalCO
    TPclass}, where the intensity of HBB changes from no-HBB to mild
  HBB and full HBB as a consequence of CBM in those stages. Also, the
  left panels in Fig.~\ref{Fig:2dup} show that third dredge-up and the
  evolution of C/O during the TP-AGB are only slightly affected by the
  occurrence of any CBM during the 2DUP event.

We conclude from this experiment that CBM during the 2DUP leaves only a
minor impact on the structure of the future TP-AGB
star, and leads to changes in 3DUP and HBB smaller
than those produced by the inclusion of CBM at other
stages of the evolution like the core H- and He-burning, and He-shell
flashes on the TP-AGB.  
                
\subsection{Impact of CBM during the TP-AGB on Third Dredge-Up Efficiency}

\label{sec: CB Single TP investigation}

An investigation performed by \cite{Herwig2000} showed the variation
in the third dredge-up efficiency\footnote{The third dredge-up
  efficiency is defined as the ratio between the mass of material
  dredged-up from the core to the mass by which the core grew during
  the preceding inter-pulse phase, given by $\lambda = {M_{\rm
      DU}}/{M_{\rm growth}}$.} $\lambda$ as a function of mixing
efficiency at the different convective boundaries. A similar approach
is taken here to determine the relative importance of the different convective
boundaries for the TDU during a single thermal pulse. Each calculation
begins with the same initial model prior to that thermal pulse, and is
evolved further with the indicated mixing efficiency applied at the
various boundaries.

\begin{figure}
  \begin{center}
     \includegraphics[width=0.8\columnwidth]{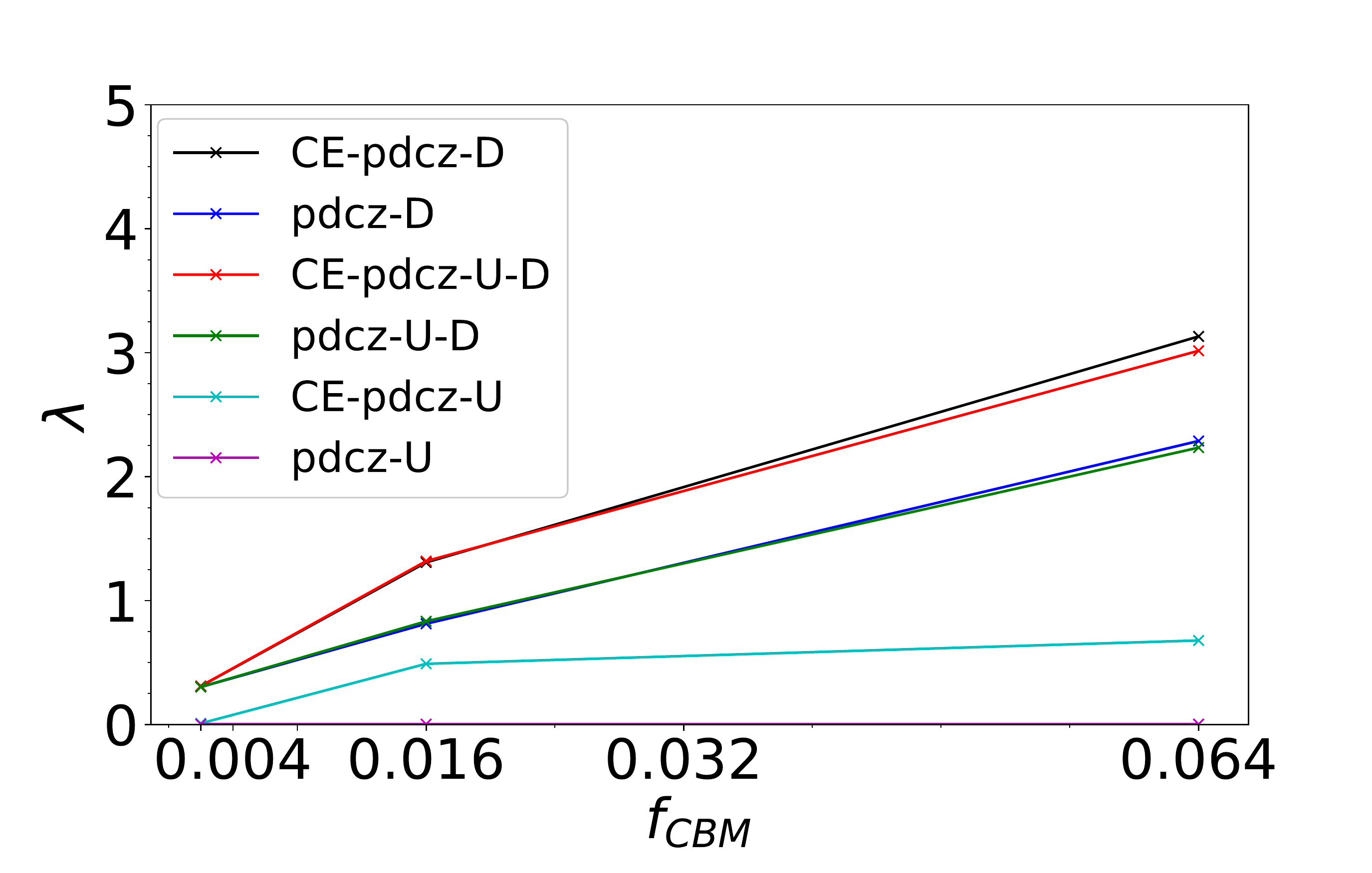}
  \caption{Dredge-up efficiency $\lambda$ as a function of convective
    boundary mixing efficiency \fCBM, for  a single thermal pulse of a 3\Msun
    model. Each line represents the inclusion of CBM at the different
    boundaries indicated. All sequences correspond to initial
      metallicities of $Z=0.02$.} 
  \label{Fig: TP11 lambda CE pdczDUB}
  \end{center}
\end{figure}

The model taken for this investigation is from a 3\Msun evolutionary
sequence just prior to the 11th thermal pulse where additional mixing
from both the hydrogen and helium burning cores was included, but no
CBM treatment during the TP-AGB was applied prior to the thermal
pulse.  It is worth determining whether both convective boundaries
  of the PDCZ, upwards (U) and downwards (D), are influential or
  not.  Fig.~\ref{Fig: TP11 lambda CE pdczDUB} shows the third
dredge-up efficiency $\lambda$ for sets of models calculated
independently of each other, with values for the parameter
\fCBM=[0.002, 0.016, 0.064] for combinations of convective boundaries
at the CE, PDCZ-U and PDCZ-D. In each case, it can be seen that mixing
upwards from the PDCZ has almost no effect, apart from the cases with
high \fCBM values in conjunction with mixing from the CE where it
alters the outcome to a small degree. This is the boundary which is
expected to have no influence on the evolution, and as this appears to
be the case in the initial testing, it is from this point on ignored.

The mixing at the base of the PDCZ influences the dredge-up through an
increase in the helium burning luminosity, a result of mixing
additional fuel from the intershell region into the helium burning
shell. This additional energy release is immediately absorbed by the
layers above the He-burning shell which consequently cool and expand.
This additional expansion and cooling of these layers, among which are
those layers at the base of the convective envelope, creates more
favourable conditions for third dredge-up as it increases the
radiative gradient, according to $\nabla_{\rm rad} \propto T^{-4}$
\citep{Herwig2000}.

\begin{figure}
  \begin{center}   
  \includegraphics[width=0.8\columnwidth]{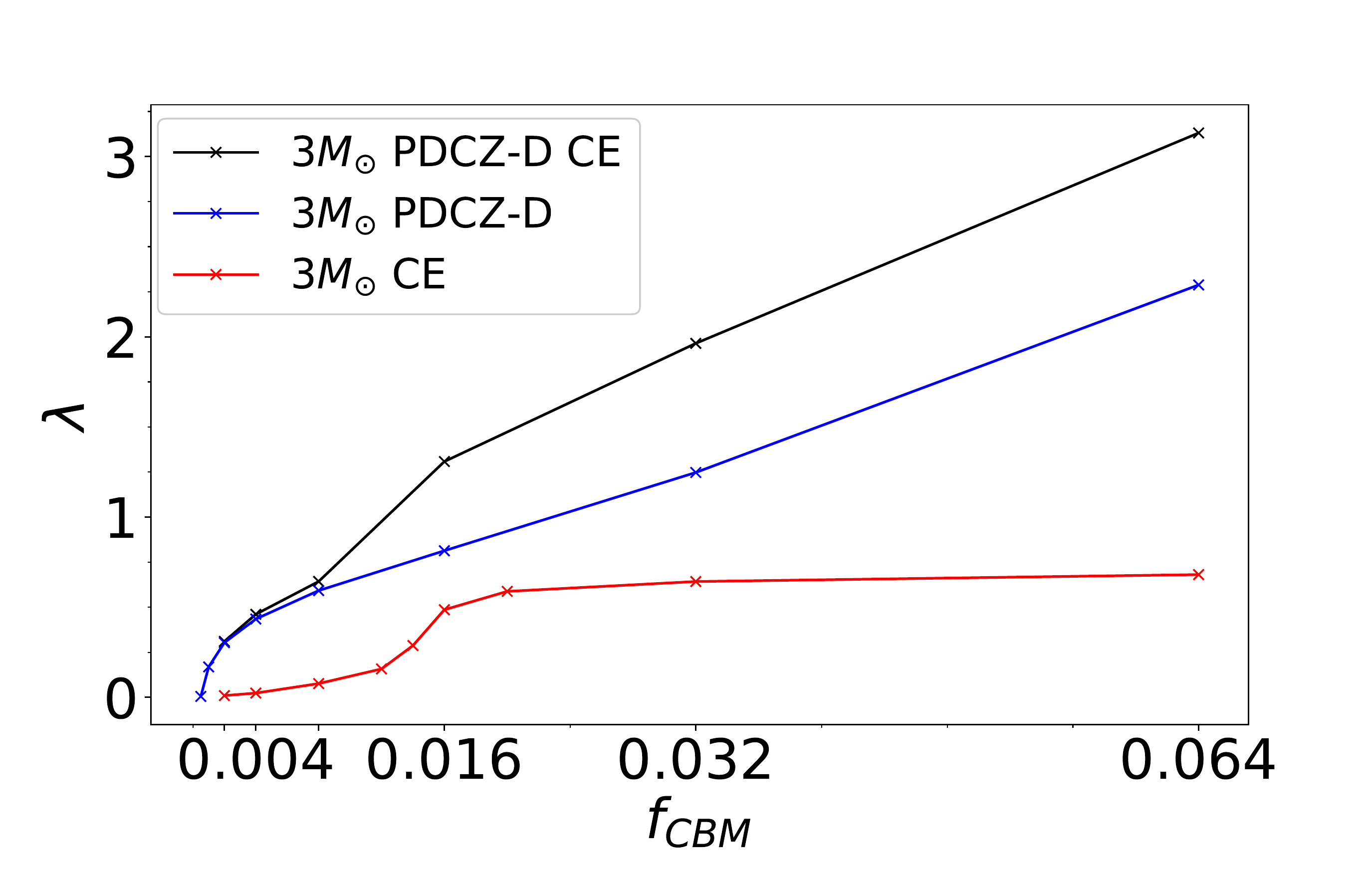}
  \caption[TDU efficiency for single thermal pulse of a 3\Msun
    model]{Dredge-up efficiency $\lambda$ as a function of convective
    boundary mixing efficiency $f_{\rm CBM}$, for a single thermal pulse of a
    3\Msun model. Each line represents the inclusion of CBM at the
    different boundaries indicated. Note that CBM was included in all
    models during previous core-burning phases.} 
  \label{Fig: TP11 lambda CE_pdczD combos}
   \end{center}
\end{figure}

A combination of the PDCZ-D and CE boundaries was then calculated for
\fCBM = [0.002, 0.004, 0.008, 0.016, 0.032, 0.064], with additional
       values for the CE-only (\fCE=[0.012, 0.014, 0.020]) and
        PDCZ-only cases (\fPDD=[0.0005, 0.001]).  The resulting third
      dredge-up efficiencies are presented as $\lambda$ in
      Fig.~\ref{Fig: TP11 lambda CE_pdczD combos}. The red line in
      this figure represents models which have \fCBM applied only to
      the CE, blue lines stand for PDCZ-D and black one for the
      combined CE-PDCZ-D.  For models where only the CE boundary is
      active, additional computations were included in order to
      resolve the sudden increase of $\lambda(f_{\rm CMB})$ at about
      \fCBM$\sim 0.016$.

 Qualitatively, there appears to be a limiting value
for $\lambda$ in the case of mixing from the convective envelope only,
as was also observed by \cite{Herwig2000}. While in this paper it
appeared as an almost ``on-off'' switch, we find a limited
range of \fCBM where the CE becomes relevant. Additionally, the
approximate limiting value of 
$\lambda \simeq 0.7$ is quantitatively comparable to that in
\cite{Herwig2000}, although this could be coincidental, given the
different initial models, methodology and operating codes.

Focusing instead on the blue line, representing the models with only
PDCZ-D, it can be seen that the base of the PDCZ already begins to
have an influence on the third dredge-up at the lower values of \fPDD
and, from \fPDD=0.004 onwards, has a linearly increasing effect on
$\lambda$ to a fairly reasonable approximation. This behaviour is also
observed in the models calculated by \cite{Herwig2000}, and has fairly
strict observational limits making it unnecessary to consider higher
\fCBM values in order to find out whether this trend could continue
indefinitely.

It is also interesting to note that the behaviour of the two
convective boundaries do not appear to be coupled, at least 
regarding their effect on the dredge-up efficiency. The black line in
Fig.~\ref{Fig: TP11 lambda CE_pdczD combos} shows the combined
CE-PDCZ-D models, which clearly represent the sum of the two
individual mechanisms. At the lower values of the CBM parameter, the
trend follows that of the PDCZ-D models, as there is no significant
influence from the CE boundary, while there is the almost step-like
increase in the CE-PDCZ-D models at the $f_{\rm CBM}=0.016$ model, to move
away from the PDCZ-D set of models before continuing with the linear
increase in $\lambda$ as the PDCZ-D boundary again entirely dominates
the outcome.

\section{Constraining Additional Mixing on the TP-AGB}
\label{sec: TP-AGB mixing}

The aim of this section is to constrain the extent of additional
mixing beyond the formal convective boundaries on the TP-AGB, through
the use of observations and the findings of the previous
section. This is primarily done through the use of carbon star
number counts, the semi-empirical initial-final mass relation (IFMR),
and the observed abundances of PG~1159 stars. It is worth emphasizing
that all sequences discussed in the following sections include CBM in
convective cores during the pre-AGB phase ($f_{\rm CHB}=f_{\rm
  CHeB}=0.0174$).

\subsection{PG~1159 Stars and intershell abundances}
\label{sec: PG1159}

The class of post-AGB stars, known as PG~1159 stars, represent the
best method of constraining the chemical composition of the intershell
region at the end of the TP-AGB lifetime of a star. The observed
oxygen abundances, in particular, cannot be explained without the
inclusion of some form of additional mixing at the base of the PDCZ,
making this particular observable a good place to start in trying to
constrain CBM in TP-AGB stars \citep{Herwig1997, Herwig2000}.

\begin{figure}
  \centering
  \includegraphics[width=\columnwidth]{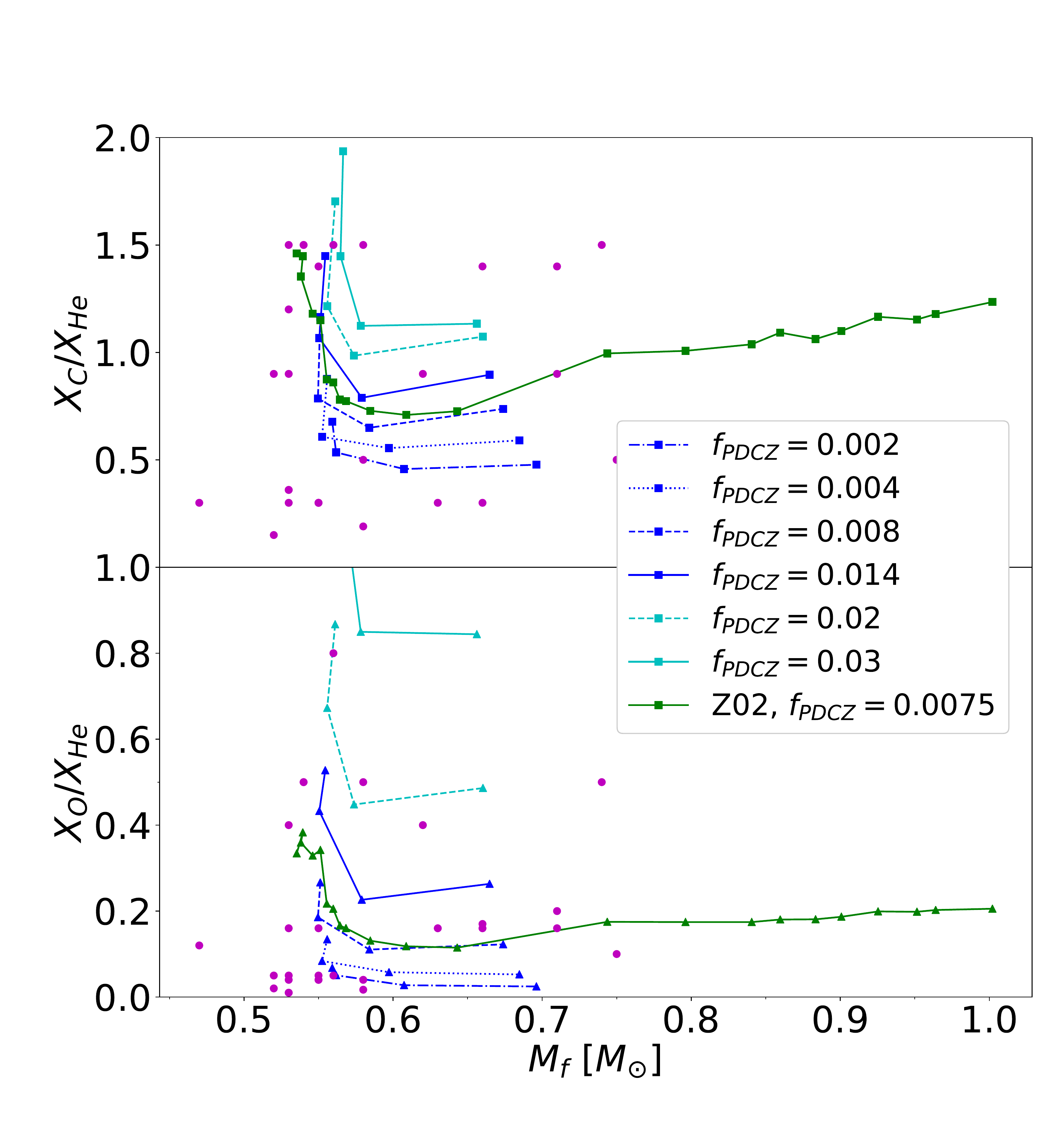}
  \caption[Intershell abundances with data]{Intershell abundances at
    the end of the TP-AGB given as the ratio of  mass fractions;
    C/He in the top panel and O/He in the lower panel, as a function
    of final mass for sets of evolutionary models with the labeled
    mixing parameter at the base of the PDCZ. ``Z02'' indicates models
    with an initial metal fraction of Z=0.02 while all others are calculated
    for Z=0.008. Purple circular markers represent
    observed abundances of PG~1159 stars and were provided by Werner (2014,
    private communication). }
  \label{Fig: IS abund PDD}
\end{figure}

The data shown in Fig.~\ref{Fig: IS abund PDD} come from the most
updated set of surface parameters $\log g$, $\log T_{\rm eff}$ and He,
C, and O abundance determinations for PG~1159 stars (Werner 2014,
private communication). Masses are derived from the comparison of
$\log g$ and $\log T_{\rm eff}$ for each object with theoretical
tracks computed by \cite{2006A&A...454..845M}.

The final intershell abundances of sets of models with varying \fPDD
at the base of the PDCZ are shown in Fig.~\ref{Fig: IS abund PDD} as a
function of final mass, along with a set of models taken from the
calculations of the previous section at Z=0.02 (denoted Z02); all
other models are for Z=0.008. For each of the Z=0.008 models, the four
initial masses are 1.6, 2.0, 2.4 and 2.8$M_\odot$, where the 1.6\Msun
model always has the highest O/He and C/He values and the 2.8\Msun
model always has the highest final mass; lines are connecting models
in the order of initial mass.   As shown in Fig.~\ref{Fig: IS
    abund PDD}, increasing \fPDD leads to an increase in the C and
  also O abundances in the intershell. In agreement with
  Fig.~\ref{Fig: TP11 lambda CE pdczDUB}, models with \fPDD $\leq
  0.004$ do not change intershell abundances significantly. Fig.  6
  shows that grids of stellar evolution models with different degrees
  of CBM at the PDCZ are able to reproduce the observed values in
  PG1159 photospheres.

The models calculated at Z=0.02 with \fPDD = 0.0075 closely match the
Z=0.008, \fPDD = 0.008 models.  Therefore having very similar CBM
  but different metallicity does not seem to impact too much on the
  predicted intershell abundances. The $Z=0.02$ models show that the
  almost constant O/He value against mass continues for all higher
  mass models calculated.  This is both reassuring in terms of a
comparable behaviour, which has been seen previously
\citep{Bertolami2016}, and that the choice of initial metallicity for
the models does not seem to impact too much on the result of the
intershell abundances.

The carbon abundances are perhaps less in favour of any particular set
of models; however, none of the sequences calculated could produce
anything like the spread of values observed there. This may be more
generally attributed to the sensitivity of the carbon abundance, which
is far more easily mixed into the intershell region.

In conclusion, for the models calculated here, a
value of \fPDD in the range 0.004-0.014 appears to be a limit of
possible CBM (in agreement with values suggested in
\citealt{Herwig2005}), with \fPDD$\sim 0.008$ being able to explain
the typical observed carbon and oxygen abundances in PG~1159
stars.  However, as the intrinsic spread of PG~1159 carbon and
  oxygen abundances are larger than errors in mass and abundance
  determinations, it seems that some object-specific form of mixing
  (such as rotation-induced mixing) might also be responsible for the
  observed abundances.

\subsection{Carbon Star Number Counts}

Rather than using the M- and C-star lifetimes as derived in
\cite{Girardi2007}, it was decided first to compare results directly to the
number counts of carbon- and oxygen-rich stars which had been gathered
for the purposes of calculating the lifetimes. This is done in order
to try and remove some of the assumptions which go into the derivation
of the lifetimes, for instance, relying on stellar isochrones having
correct ages for all phases of stellar evolution. Of course, errors
due to the stellar isochrones still enter into determinations of the
initial masses of the objects in a cluster.

The clusters in the sample of \cite{Girardi2007} were binned according
to the turnoff mass of the cluster for the LMC and SMC
separately. Although this represents an excellent sample in terms of
AGB stars, the numbers are still very small and so the focus presented
here is on three of the data bins taken from their Table~2 and which
are summarised in Table~\ref{Tab: LMC nCounts}. These points are
representative of clusters in the LMC, which have the highest numbers
of both M- and C-stars, and also represent the mass range where carbon
stars are expected from the stellar evolution models at this
metallicity.

\begin{table}
\centering
  {\renewcommand{\arraystretch}{1.3}
  \begin{tabular}{l |  c  c c| c c c}
    \hline
    \hline
    $M_{\rm TO}$ & $N_{\rm M}$ & $N_{\rm M}^{\rm min}$  & $N_{\rm M}^{\rm max}$  & $N_{\rm C}$ & $N_{\rm C}^{\rm min}$  & $N_{\rm c}^{\rm max}$   \\    \hline
    1.66 & 9  & 5 & 12 & 10 & 6 & 13 \\
    2.17 & 22 & 16& 27 & 32 & 25& 38 \\
    2.66 & 4  & 1 & 6  & 4  & 1 & 6 \\
     \hline
  \end{tabular}
  }
\caption[Turnoff mass and AGB number counts]{Turnoff mass ($M_{\rm
    TO}$) along with the number of M-stars ($N_{\rm M}$) and C-stars
  ($N_{\rm C}$) taken from Table~2 in \cite{Girardi2007} for binned
  LMC cluster data. The minimum and maximum values taken for a given
  observation of N objects, were calculated as discussed in the text.} 
\label{Tab: LMC nCounts}
\end{table}

The number counts allow for direct comparison with the stellar
evolution models, as it is simple to take the time spent as either
oxygen-rich or carbon-rich for any given evolutionary track. There is
the additional minimum luminosity cutoff of $L_{\rm
  cutoff}$=3.336$L_{\odot}$, corresponding to $M_{\rm bol}$=-3.6 which
was used as an observational limit for ensuring the stars observed
were AGB stars above the tip of the RGB. Thus the fraction of the
total number of observed stars which are carbon-rich can be compared
to the stellar models. Additionally, to give some indication of how
much weight each point should carry, simple upper and lower limits are
included in all figures. These limits are taken from considering a
Poisson distribution for each number count and taking the maximum and
minimum expected number count which would correspond to covering 68\%
of the distribution. For a given data point, there is both an M-star
and C-star number count, so if the minimum M-star count possible
(within this very wide tolerance) and maximum C-star count are taken,
this should give the maximum observed C-star fraction possible within
these limitations, i.e. the maximum C-star fraction is given by
\begin{eqnarray}
\left(\frac{N_{\rm C}}{N_{\rm Tot}}\right)^{\rm max} = \frac{N_{\rm C}^{\rm max}}{N_{\rm C}^{\rm max}+N_{\rm M}^{\rm min}}.
\end{eqnarray}

Equally, the minimum C-star fraction can be obtained by taking the
maximum value $N_M^{\rm max}$ and the minimum value $N_C^{\rm
  min}$. The minimum and maximum values for the number counts taken
were as listed in Table \ref{Tab: LMC nCounts}.  It must be stressed
that this should not be interpreted as a claim of statistical
significance, and was done in the interests of indicating the
respective importance of each data point.

\label{Sec: Obs Ev  Mod Star Counts}

\begin{figure}
  \centering
  \includegraphics[width=\columnwidth]{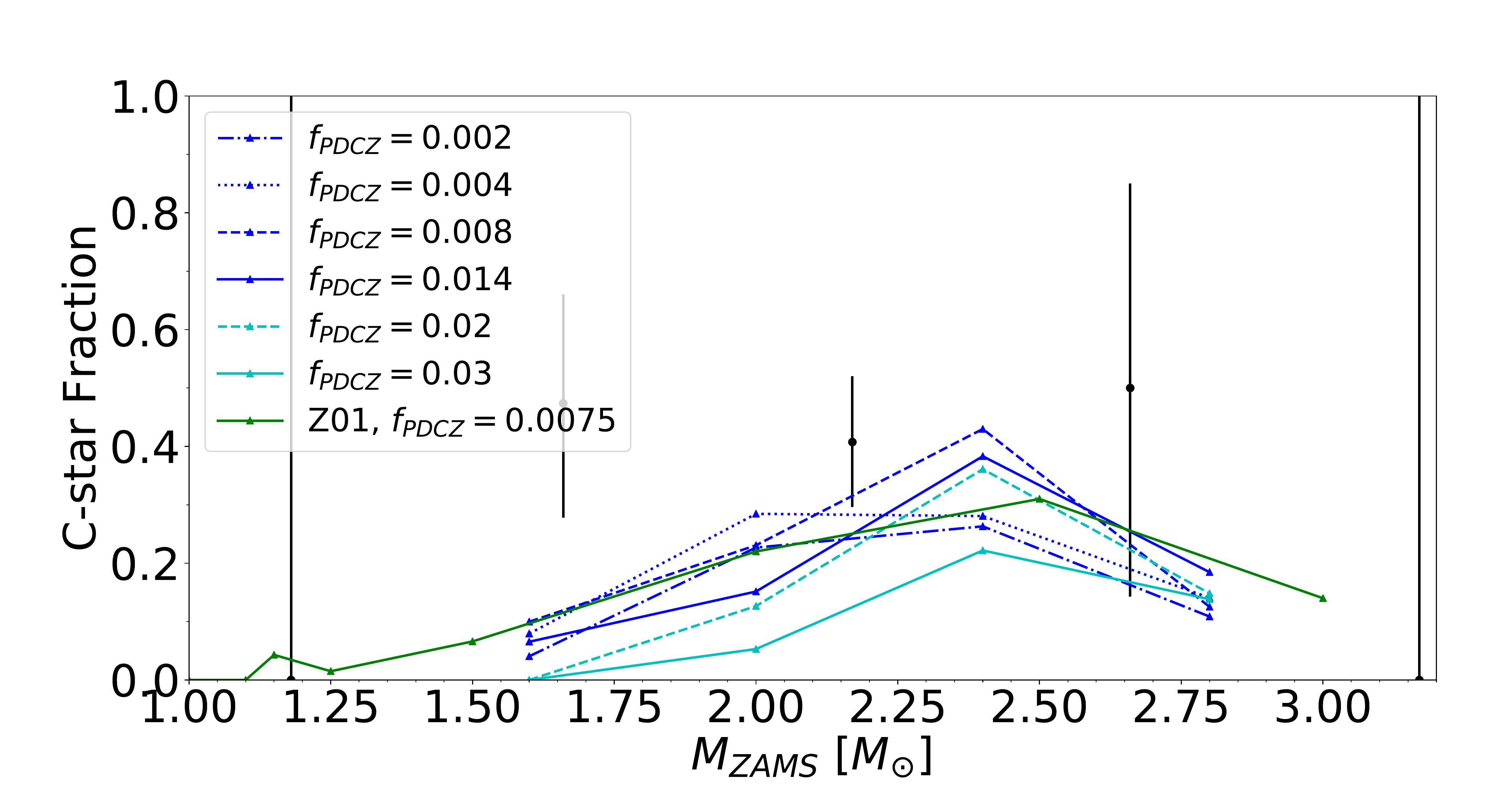}
  \caption[C-star fractions for PDCZ CBM models]{Carbon star fractions
    for models where CBM is applied only at the base of the PDCZ
    during the TP-AGB phase, all calculated at Z=0.008, with data
    points from Table 2. The $Z=0.01$, \fPDD=0.0075 models are taken
    from  \cite{Bertolami2016}. C-star fractions for the Large
    Magellanic Cloud are taken from \cite{Girardi2007} as discussed in
    the text.} 
  \label{Fig: cFrac PDD}
\end{figure}

Models where CBM only at the base of the PDCZ is active are shown in
Fig.~\ref{Fig: cFrac PDD} in terms of their carbon star fraction. The
range of \fPDD covers 0.002-0.03, and the first point to note is that
the C-star fraction does not continue to increase with increasing
\fPDD as may have been expected from the results in \S \ref{sec: CB
  Single TP investigation}.  Models with \fPDCZ in the range
  $0.004$ and $0.02$ lead to the largest C-star fractions in
  Fig.~\ref{Fig: cFrac PDD}, no model appears to satisfactorily
  reproduce the C-star fractions, even accounting for binning and
the coarse grid of models calculated. On the whole, the impression is
that additional mixing at the base of the PDCZ alone is not sufficient
to reproduce the observed numbers of carbon stars, for any value of
$f_{\rm PDCZ}$.  Still, as we will see in section \ref{subsec:tension}
  a larger C-star fraction can be easily reproduced by a decrease in
  the C-rich wind prescription adopted in the models.

However, it was seen in the single thermal pulse investigated in \S
\ref{sec: CB Single TP investigation} that increasing the mixing
parameter at the base of the PDCZ would lead to a strong, almost
linear increase in the dredge-up efficiency. Fig.~\ref{Fig: lamMax
  PDD} shows the maximum dredge-up efficiency parameter, $\lambda_{\rm
  max}$, as a function of initial mass for the models where only the
base of the PDCZ is active, the behaviour clearly does follow that
expected from the single thermal pulse investigation in \S \ref{sec:
  CB Single TP investigation}.
\begin{figure}
  \centering
  \includegraphics[width=\columnwidth]{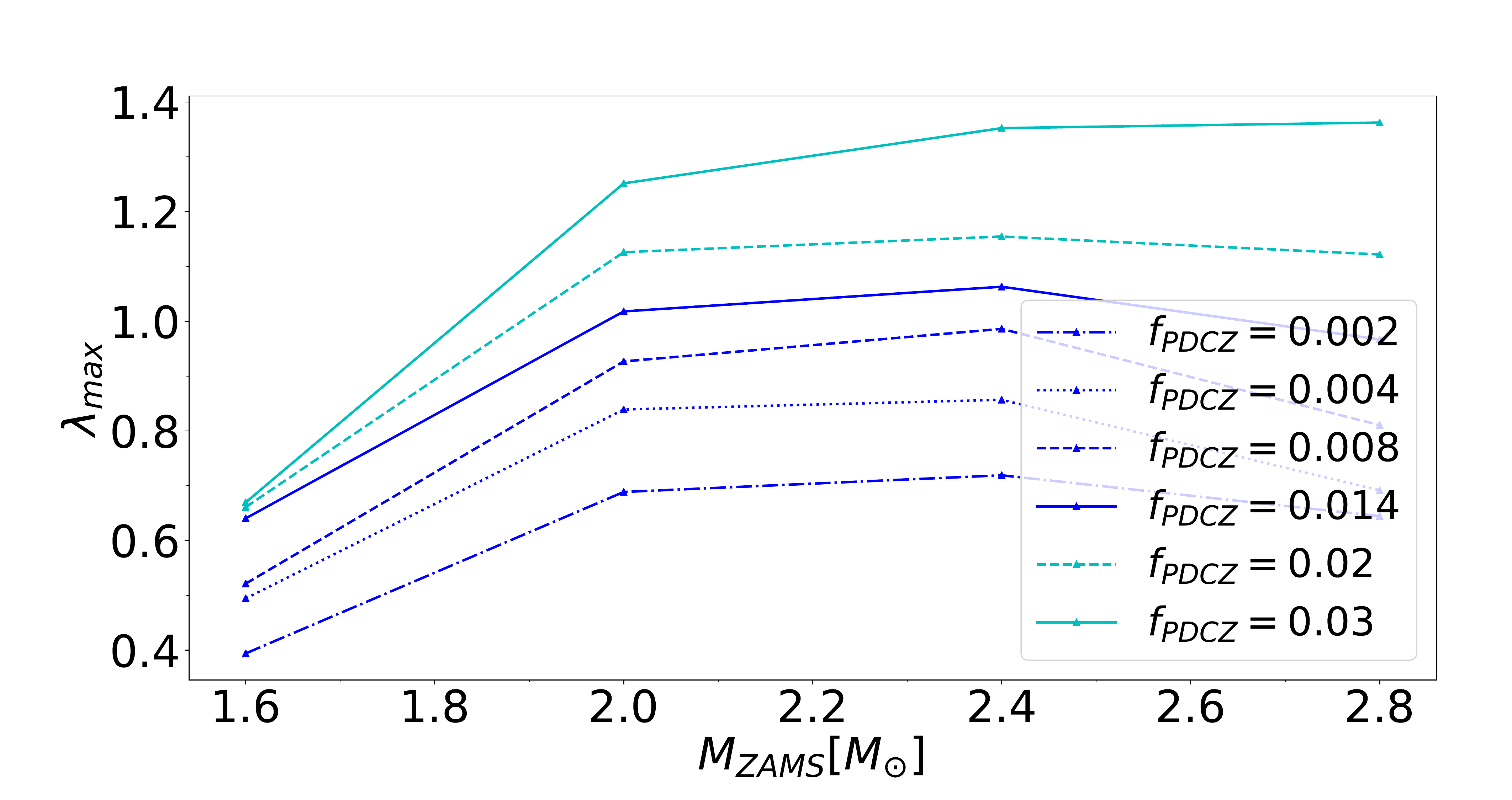}
  \caption[TDU efficiency vs initial mass]{Maximum value for the third
    dredge-up efficiency parameter, $\lambda$, experienced by each
    model sequence under the assumption of different values of \fPDD.} 
  \label{Fig: lamMax PDD}
\end{figure}
 Although higher values for \fPDD do indeed lead to more efficient
 dredge-up, stronger mixing at the base of the PDCZ also changes the
 composition of the intershell (Fig.~\ref{Fig: IS abund PDD}),
 something which is primarily determined during the early thermal
 pulses as the intershell forms, only changing slightly
 thereafter. For example, for the 2.4\Msun, a value of \fPDD=0.03
 leads to a C to O ratio of C/O=1.36, compared to a value of C/O=3.51
 for \fPDD=0.008.  Consequently, the material dredged up to the
 surface has a higher O content in the case of a large value of \fPDD
 delaying the increase of the C/O ratio and the transition to a
 C-star, thus reducing the C-star fraction predicted by the models
 (see Fig.~\ref{Fig: cFrac PDD}). Another consequence of the higher O
 content of the intershell material when \fPDD is increased, can be
 seen in the final surface C/O ratios of the models. While the
 2.4\Msun with \fPDD=0.03 ends the TP-AGB evolution with a surface
 value of C/O=1.3, the 2.4\Msun with \fPDD=0.008 ends the TP-AGB with
 a much higher surface C/O ratio of C=3.7.

\begin{figure}
  \centering
  \includegraphics[width=\columnwidth]{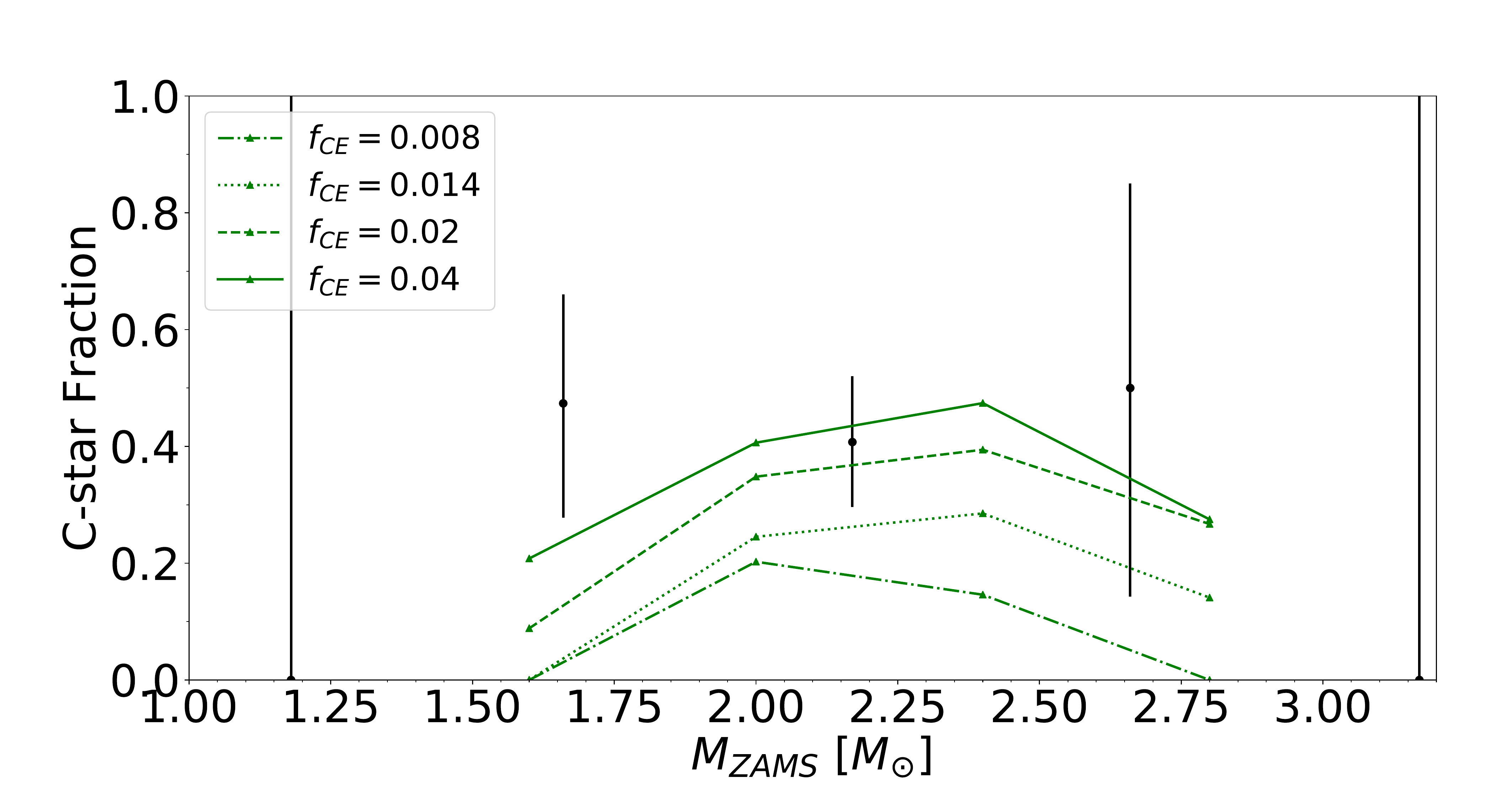}
  \caption[C-star fraction for CE CBM models]{Carbon star fractions
    for models where CBM is applied only at the base of the convective
    envelope, all calculated at Z=0.008, with data points from Table~2
    of \cite{Girardi2007} and error bars as discussed in the text.} 
  \label{Fig: cFrac CE}
\end{figure}

Fig.~\ref{Fig: cFrac CE} shows a similar set of models as
Fig.~\ref{Fig: cFrac PDD}, this time displaying the C-star fractions
for models with varying \fCE between 0.008 and 0.4 at the base of the
convective envelope. In this instance, there is a far more
straightforward interpretation, with an increase in \fCE producing an
increased C-star fraction, in almost all cases. Furthermore, there is
more readily acceptable agreement with the observed C-star fractions
than was the case for the models including only $f_{\rm PDCZ}$. The data point
at the lowest mass is perhaps still too high, though it is not
unreasonable to think the models are sufficiently close when binning
and a better-resolved grid of models are taken into account.
\begin{figure}
  \centering
  \includegraphics[width=\columnwidth]{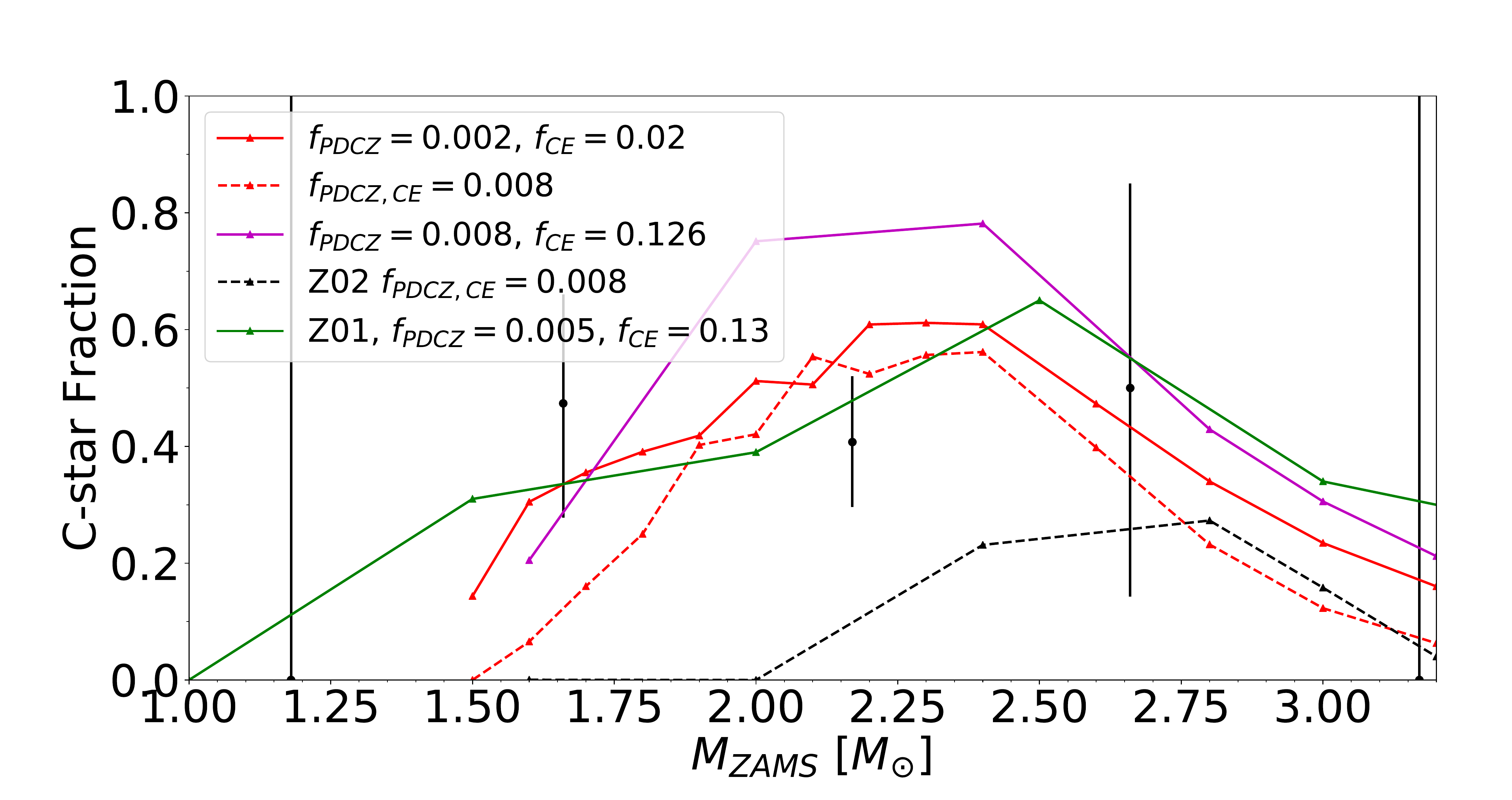}
  \caption[C-star fractions for PDCZ CE CBM models]{Carbon star
    fractions for models where CBM is applied at both the base of the
    PDCZ and the convective envelope, all calculated at Z=0.008 except
    for the black dashed line at Z=0.02, with data points from Table~2
    of \cite{Girardi2007} and errors as discussed in the text.} 
  \label{Fig: cFrac PDD CE}
\end{figure}
Models were also produced for a range of combinations of \fPDD and
\fCE in conjunction, with a focus on models where the two are
equal. Additionally, models where the value of \fCE is higher than
\fPDD were produced following the results from \S \ref{sec: Max OS}
where a large discrepancy in the allowed momentum-based CBM at the
respective boundaries was suggested. Finally, the preferred value for
mixing at the convective envelope, which is derived for the production
of the \Iso{13}{C} \citep{Herwig2003}, was also considered, namely
\fCE=0.126, which is higher than would have otherwise been
applied. This is sometimes taken alongside a value of \fPDD=0.008
\citep{Pignatari2016,Ritter2017} for the reproduction PG~1159 star
abundances.
A selection of these models, considered to be of interest, are
presented in Fig.~\ref{Fig: cFrac PDD CE}. The first set of models
(magenta line) are those with \fPDD= 0.008, \fCE= 0.126 and which
produce C-star fractions close to 80\% at both 2.0\Msun and
2.4$M_\odot$. Even with the possible influence of binning of the
observed data points, and the fact that these two masses do not
represent the peak C-star fraction (as exhibited by the other models),
this seems to give a first indication that the values sometimes taken
for CBM on the TP-AGB \citep{Pignatari2016,Ritter2017} do not
necessarily agree with evolutionary observational evidence, even if
they are well motivated by other means.  The black line corresponds to
models with Z=0.02, and a choice of \fPDD=\fCE= 0.008 to show the
influence of metallicity, when compared with the other models with the
same choice of \fPDD and \fCE. Not only are the C-star fractions
significantly lower; in this case, the peak also shifts to higher
initial masses, emphasising that composition plays a significant role
in how the models relate to the observations.  The other lines
  correspond to sets of models with (\fPDD,\fCE) = (0.008,0.008) and
  (\fPDD,\fCE) = (0.002,0.02), as well as the {\tt LPCODE} model with
  (\fPDD,\fCE) = (0.005,0.13) which are indeed able to reproduce the
  observed carbon star fractions. Yet a major inconvenience arises when
  absolute lifetimes of M-stars and C-stars are considered, as shown
  in Fig.~\ref{Fig: cLife PDD CE}.

\begin{figure}
  \centering
    \includegraphics[width=\columnwidth]{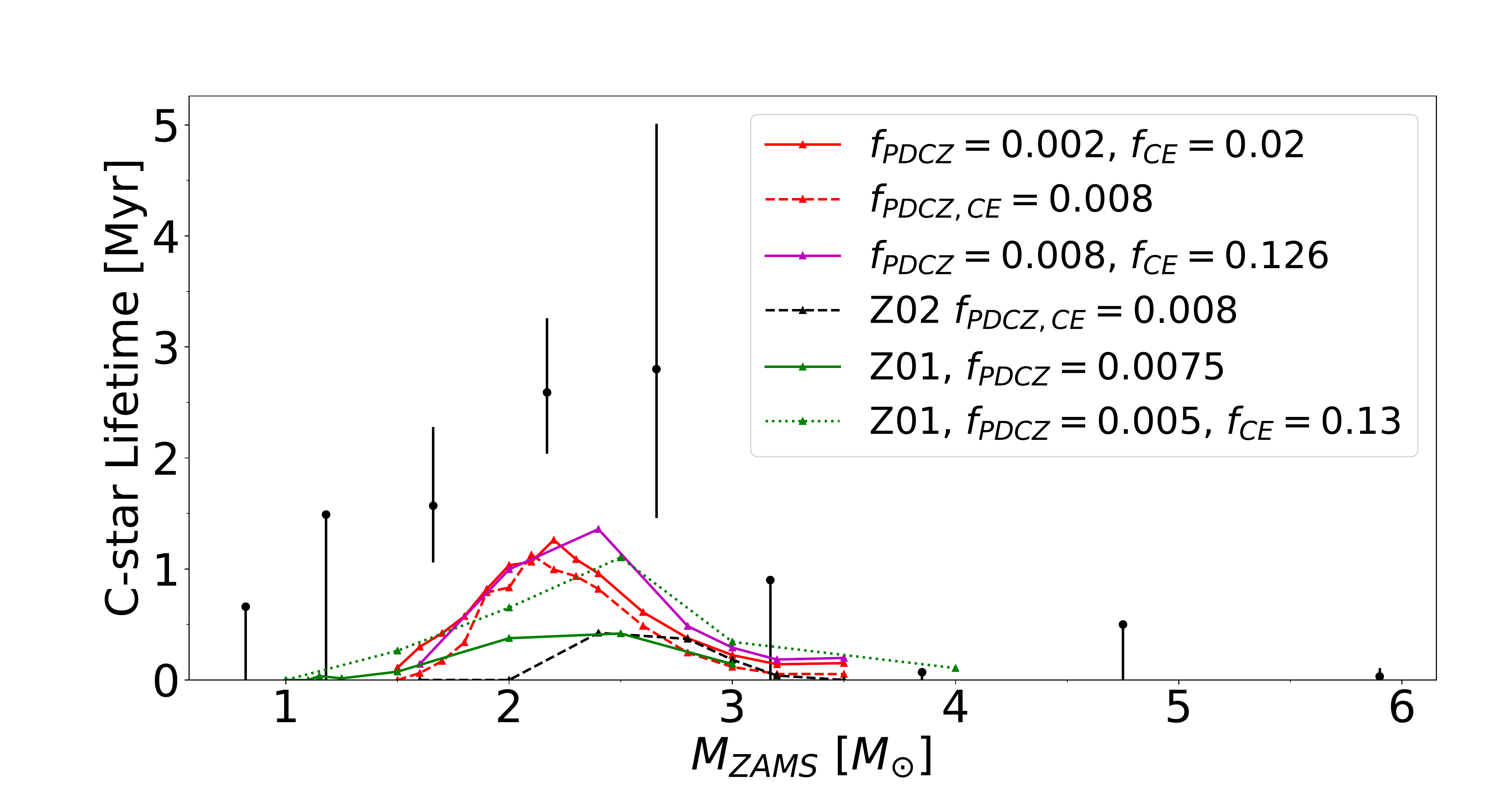}
  \caption[C-star lifetimes CBM models]{Carbon star lifetimes for
    the same models as in Fig.~\ref{Fig: cFrac PDD CE}.
    In addition, an additional case, computed with {\tt LPCODE}, and
    parameters $Z=0.01$, $f_{\rm PDCZ}=0.005$ is shown (solid green
    line). Data points are derived from \cite{Girardi2007}.} 
  \label{Fig: cLife PDD CE}
\end{figure}

As an additional comparison, the C-star lifetimes of the same sets of
models as in Fig.~\ref{Fig: cFrac PDD CE} are shown again in
Fig.~\ref{Fig: cLife PDD CE}, in order to directly compare with the
data markers derived from the stellar number counts in
\cite{Girardi2007}. Although all the models calculated at Z=0.008
reproduce the basic shape of the observations, the lifetimes are 2-3
times lower than the derived lifetimes, even in the case with
(\fPDD,\fCE) = (0.008,0.126).  A similar result was found in LPCODE
models by \cite{Bertolami2016}. Again, the set of models calculated
for Z=0.02 demonstrate how important the initial composition can be in
such comparisons. In fact, none of the models calculated here comes
close to reproducing the C-star lifetimes. Previously, models
calculated with {\tt GARSTEC} \citep{Kitsikis2008,Weiss2009} were in
better agreement with these observables. This was in fact due to the
individual models calculated at 2\Msun experiencing the anomalous
phenomenon reported in \cite{Wagstaff2018}, which increases the C-star
lifetime around an initial mass of 2\Msun in the models of
\cite{Wagstaff2018}.  Additionally, the discrepancy between
  observationally inferred C- and M-star lifetimes by
  \cite{Girardi2007} and those produced by stellar evolution models
  might also be related to the AGB boosting effect described by
  \cite{2013ApJ...777..142G} for clusters around 1.6 Gyr. As it will
  be discussed in Section \ref{subsec:tension}, this effect might be
  affecting the derived lifetimes in the 1.66$M_\odot$ and 2.17
  $M_\odot$ LMC bins of \cite{Girardi2007} (Fig.~\ref{Fig: cLife PDD
    CE}) by up to a factor of two (Marigo \& Girardi, 2019, private
  communication). Thus signifficantly reducing the discrepancy with
  the models.

\subsubsection{IFMR}

The method of using white dwarfs in open clusters to construct a
semi-empirical IFMR is a very useful tool for constraining the TP-AGB
evolution. Examples of this method are \cite{Salaris2009},
\cite{Kalirai2014}, and \cite{2018ApJ...866...21C}. In addition,
\cite{2018ApJ...860L..17E} suggested a new method using GAIA data
which might help to constrain the low-mass limit of the IFMR.

\begin{figure}
  \centering
    \includegraphics[width=\columnwidth]{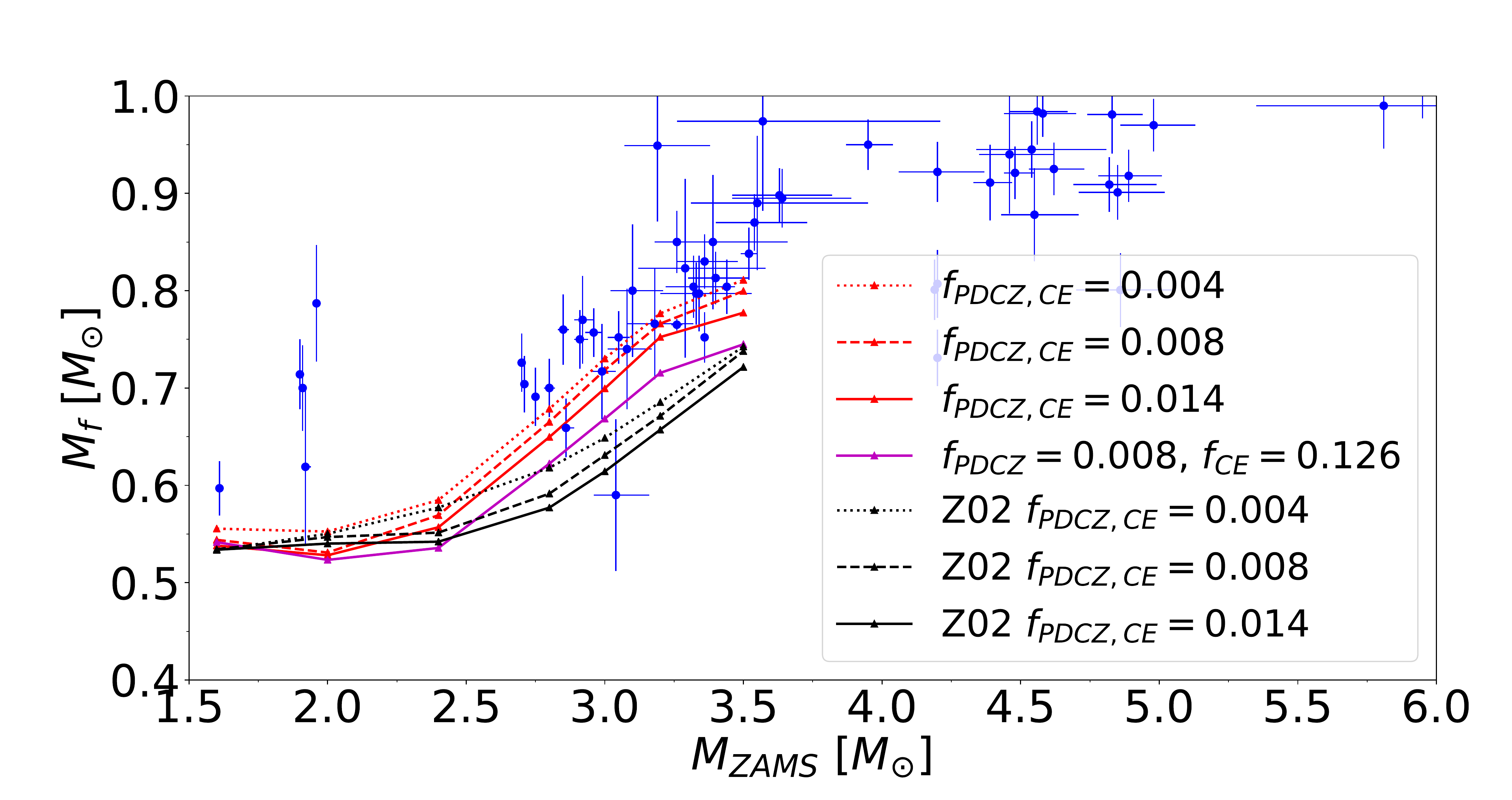}
  \caption[IFMR for CBM models]{Initial-final mass relation for models
    where CBM has been applied at both boundaries for several values,
    black lines are all calculated at Z=0.02, all others at
    Z=0.008. Data points are from the semi-empirical IFMR of
    \cite{2018ApJ...866...21C}. } 
  \label{Fig: IFMR PDD CE}
\end{figure}

The IFMR is shown in Fig.~\ref{Fig: IFMR PDD CE}, where the red lines
are models where \fPDD= \fCE for different values at Z=0.008, while the
black lines are their counterparts calculated at Z=0.02. The models
with \fPDD= 0.008, \fCE= 0.126 calculated at Z=0.008 are also
included. The models with \fPDD= \fCE at Z=0.008 all appear to be in
relatively good agreement with the main clump of observed objects
around 3\Msun and, interestingly, the change arising from the
different CBM treatments is significantly less than that arising from
the change in initial composition. The models calculated at Z=0.02
were intended to be the 'solar' case for the calculations performed,
and hence would allow for reasonable comparison with the
IFMR. Although this is likely to be too high a metallicity for solar,
it should still be the closest of the models calculated and it would
have to be said that none of these models give particularly good
agreement with the 3\Msun observations.

Additionally, at the lower mass end ($\lesssim 2M_{\odot}$) the models
are also at the lower boundary of agreement with the final masses
observed.  That all sets of models calculated here roughly agree on
the final mass of the stars at the lower mass end, and not with
observations, suggest something else may be the problem, with mass
loss being the most obvious candidate. However, as seen in the
previous section, CBM in the pre-AGB phase can affect the mass of the
H-free core at the moment of the first thermal pulse and, in that way
alter the final mass of the white dwarf.

The models calculated for (\fPDD, $f_{\rm CE}$) = (0.008, 0.126) are again shown
for comparison. They also appear to be in tension with the
observations, even though these models are calculated at Z=0.008. The discrepancy would only get worse if these were also calculated at
a higher metallicity of $Z=0.02$.

\begin{figure}
  \centering
  \includegraphics[width=\columnwidth]{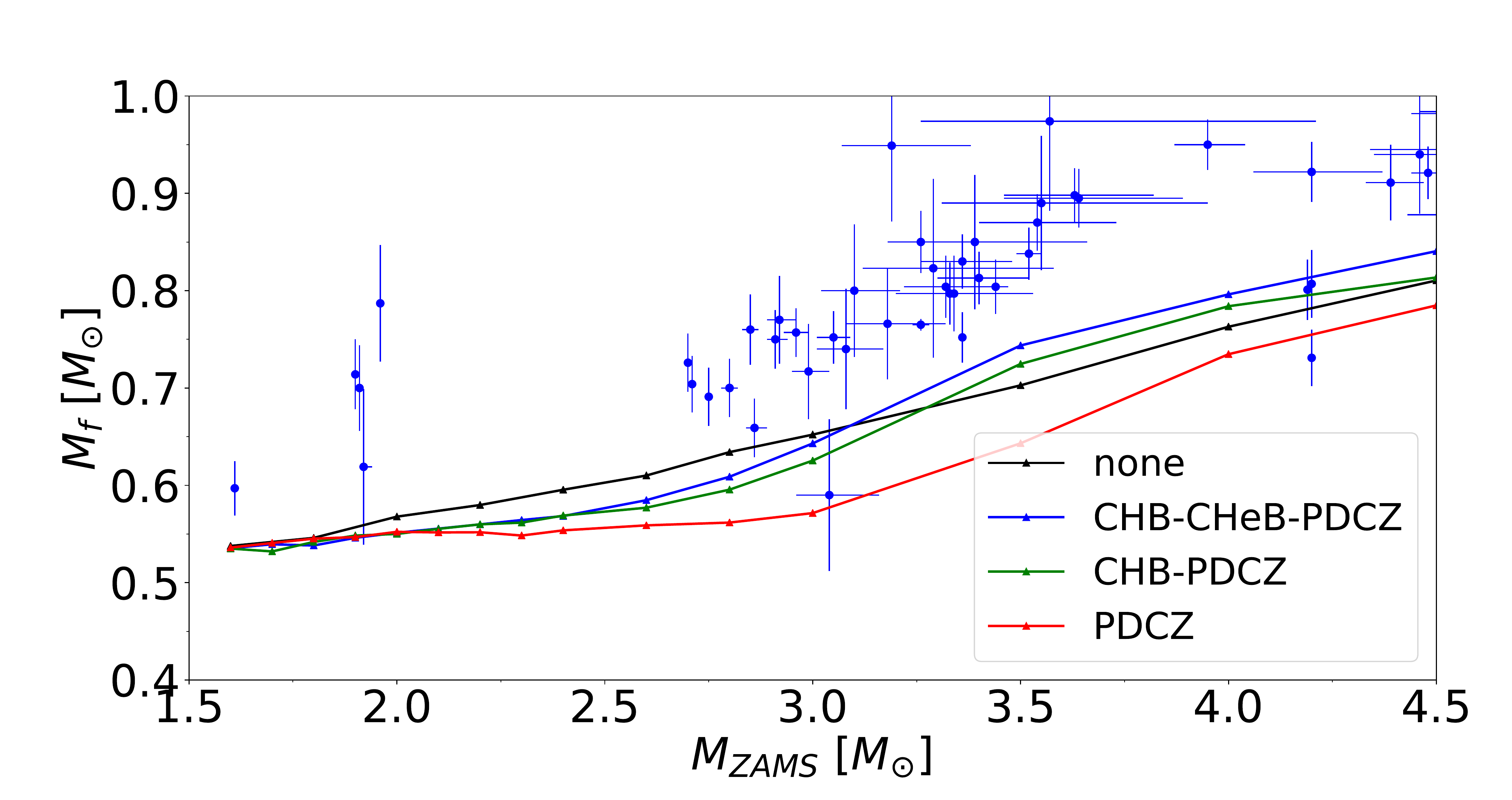}
  \caption[IFMR for core CBM models]{Initial-final mass relation for
    models where CBM has been applied during different convective core
    stages and at the base of the PDCZ for all except the black
    line. All are calculated at Z=0.02. Data points are as in
    Fig.~\ref{Fig: IFMR PDD CE}.} 
  \label{Fig: IFMR coreOS}
\end{figure}

For interest, the models from \S \ref{sec: CB EvCalc Core} studying
the influence of CBM during phases prior to the TP-AGB are shown in
Fig.~\ref{Fig: IFMR coreOS}. As these models were all calculated with
Z=0.02, it is not surprising that they are below the observational
markers in the 3\Msun region. The inclusion of core CBM does increase
the final masses of the stars, and possibly lead to a functional form
which better represents the data, but the final masses still appear to
be too low, even though for these models \fCE=0 and \fPDCZ=0.0075
(apart from the model with no additional mixing).   We can
  conclude that while CBM in pre-AGB stages is in agreement with the IFMR, CBM
  during the AGB (and in particular at the PDCZ) is clearly disfavoured
  by the IFMR.

\subsection{Tension between different observational constraints}
\label{subsec:tension}

\begin{figure}
  \centering
  \includegraphics[width=\columnwidth]{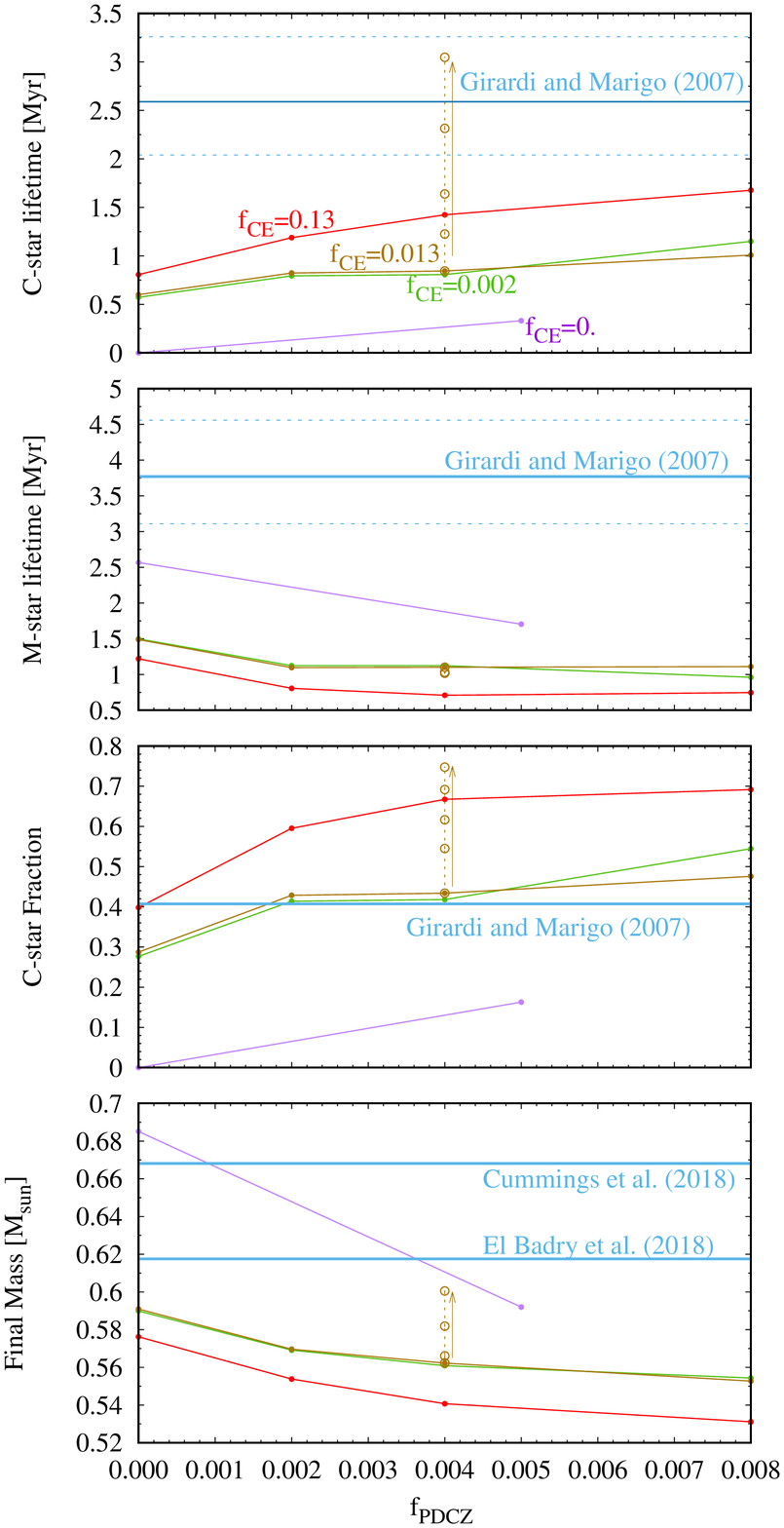}
  \caption[Multi]{Dependence of different properties of TP-AGB stellar
    evolution models with $M_i=2.2$\Msun on the values of the CBM
    intensity at the PDCZ ($f_{\rm PDCZ}$) and convective envelope
    ($f_{\rm CE}$) as compared with observationally derived values
    from clusters in the Large Magellanic Cloud \citep{Girardi2007}
    and the semi-empirical IFMR of galactic open clusters
    \citep{2018ApJ...866...21C,2018ApJ...860L..17E}.  Open symbols
      correspond to the sequences ($f_{\rm CE}=0.013$, and $f_{\rm
        PDCZ}=0.004$) computed with a reduction of the mass loss rate
      by a factor 2, 4, 8, and 16 shown in
      Fig. \ref{Fig:Multi_Mdot}. The arrows indicate the how values
      change as mass loss rates are decreased.  }
  \label{Fig:Multi}
\end{figure}

From the comparison of our model intershell abundances and the
observed O-abundances of PG1159 stars it is clear that CBM at the PDCZ
with a value of $0.004\lesssim f_{\rm PDCZ}\lesssim 0.014$ is to be
preferred (see Fig.~\ref{Fig: IS abund PDD}). This is in agreement
with previous findings by \cite{Bertolami2016, Pignatari2016,
  Ritter2017}. However, when the TP-AGB C-star fraction derived from
the LMC is considered \citep{Girardi2007}, the inclusion of CBM only
at the PDCZ gives C-star fractions, which are systematically too low in
the range $1.50 M_\odot\lesssim M_i \lesssim 2.50 M_\odot$ (see
Fig.~\ref{Fig: cFrac PDD}). The agreement with the carbon star
fraction is much improved when CBM is included also at the convective
envelope (Fig.~\ref{Fig: cFrac CE}). The agreement is particular good
for the sets with ($f_{\rm PDCZ}$, $f_{\rm CE}$) = (0.002, 0.02) in
case of the {\tt GARSTEC} code and ($f_{\rm PDCZ}$,$f_{\rm CE}$)=
(0.005, 0.13) for the {\tt LPCODE}. The need for CBM at the bottom of
the CE is in agreement with the requirements from s-process
nucleosynthesis, which suggests values of the order $f_{\rm
  CE}=0.1...0.2$ during third dredge-up episodes
\citep{Cristallo2009, Ritter2017}.

This emerging picture of moderate CBM at the bottom of the PDCZ and a
larger value at the bottom of the CE is in agreement with our previous
estimation of the possibility of turbulent entrainment at both
convective boundaries during thermal pulses (Section \ref{sec:
  RiB}). However, despite this nice agreement, this picture fails
miserably to reproduce other observables, such as the total lifetimes
of TP-AGB M-stars and the shape of the IFMR.

In fact, when convection is included both at the CE and at the PDCZ,
the final masses predicted by stellar evolution sequences fall well
below the range indicated by semi-empirical IFMR
\citep{2018ApJ...866...21C,2018ApJ...860L..17E}. This suggests that
either the TP-AGB is too short or that third dredge-up is too strong,
inhibiting the growth of the H-free core during the TP-AGB. A similar
result was obtained by \citet{Kalirai2014} who found, by
calibrating third dredge-up efficiency in TP-AGB envelope models,
third dredge-up to be too strong in full stellar evolution
models. This conclusion is in agreement with the indications coming
from the absolute lifetimes of C- and M-stars predicted by these model
sequences --see Figs. \ref{Fig: cLife PDD CE}, \ref{Fig:Multi} and also
\cite{Bertolami2016}. 

Fig.~\ref{Fig:Multi} highlights these tensions between different
observables in the case of a $M_i=2.2 M_\odot$ sequence (which is the
most unfavourable case). In order to reproduce the expected final
masses from semi-empirical IFMR and TP-AGB lifetimes (in particular
M-stars) models would need to have no CBM at neither the PDCZ nor the
CE\footnote{Also, O-rich winds on the TP-AGB would need to be
  reduced.}. On the contrary, in order to reproduce the C-star
fraction, PG1159 abundances and s-process nucleosynthesis, CBM is
needed at both convective boundaries. This is in line with the
conclusions from the recent study by
\cite{Pastorelli,2019MNRAS.485.5666P} on the basis of evolutionary
TP-AGB envelope models.

\begin{figure}
  \centering
  \includegraphics[width=\columnwidth]{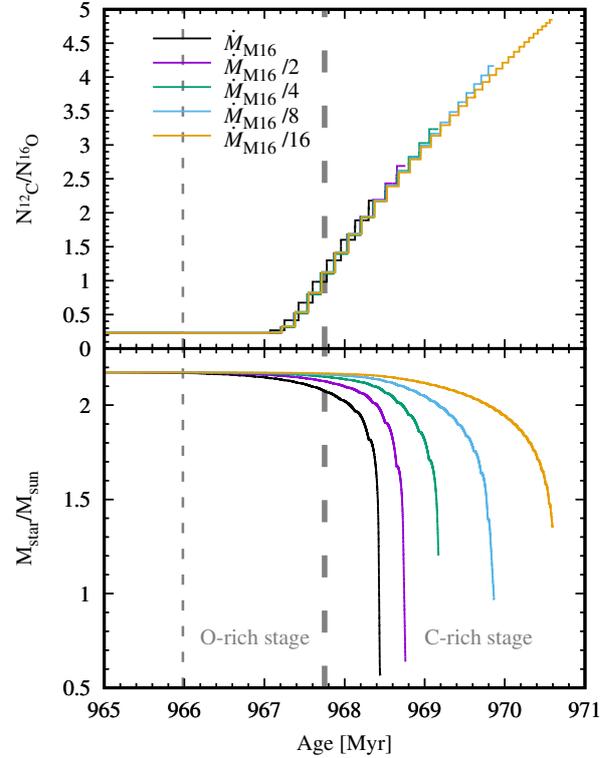}
  \caption[Multi]{ Evolution of the $M_{\rm ZAMS}=2.2 M_\odot$ model
    ($f_{\rm CE}=0.013$, and $f_{\rm PDCZ}=0.004$, see
    Fig. \ref{Fig:Multi}) during the TP-AGB under the assumption  of
    different mass loss intensities. Black, purple, green, blue, and
    orange lines correspond to sequences computed with
    $\dot{M}=\dot{M}_{\rm M16}$, $\dot{M}_{\rm M16}/2$,  $\dot{M}_{\rm
      M16}/4$, $\dot{M}_{\rm M16}/8$, and $\dot{M}_{\rm M16}/16$ where
    $\dot{M}_{\rm M16}$ corresponds to the TP-AGB wind recipes adopted
    by \cite{Bertolami2016}. The upper panel shows evolution  of
    the C/O ratio against time, while the lower panel shows that of
    the stellar mass. Vertical thin and thick dashed 
    lines indicate the beginning of the TP-AGB, and the point at which
    the models transform from an O-rich to an C-rich surface
    composition, respectively.} 
  \label{Fig:Multi_Mdot}
\end{figure}

 It is important to note that the well-known uncertainties in the
 strength of TP-AGB winds do not alter this conclusion, as they do not
 reduce these tensions. In fact, while the long C-stars lifetimes
 derived by \citep{Girardi2007} can indeed be accommodated by
   reducing C-rich winds on the TP-AGB models (see Figs
   \ref{Fig:Multi} and \ref{Fig:Multi_Mdot}), the situation is very
 different for the length of the M-star phase of the same
   sequences.  Contrary to what happens in lower mass AGB stars
   that do not experience 3DUP, or in more massive stars where HBB
   keeps the surface composition O-rich, the length of the M-star
   phase of stellar models that form C-rich stars is almost
   exclusively determined by the intensity of 3DUP. This is because it
   is 3DUP that dominates the timing at which the surface composition
   is transformed from an O-dominated atmosphere (C/O<1) to a
   C-dominated one (C/O>1), ending the M-star phase (see
   Fig. \ref{Fig:Multi_Mdot}). Such transformation depends on the
 amount of C dredged-up to the surface, and -- in light of previous
 discussions -- on the assumptions about CBM at the bottom of the CE
 and the PDCZ, but not on the intensity of winds. The only role that
 winds can play in this context is to reduce the H-rich envelope, such
 that a given mass of carbon dredged-up to the surface leads to a
 larger enhancement of photospheric C due to a lower dilution. Such an
 effect would shorten the O-rich TP-AGB phase, instead of lengthening
 it. In fact, given that standard mass loss prescriptions imply a
 rather small erosion of the envelope during the O-rich phase (see
 Fig. \ref{Fig:Multi_Mdot}) our estimations of the M-star lifetime can
 be understood as the upper limit for each given assumption of
 CBM. The fact that mass-loss intensity does not affect the conclusion
 of the previous paragraphs can be appreciated in
 Fig.~\ref{Fig:Multi_Mdot}, where we show the result of varying the
 intensities of winds ($\dot{M}_\star$) for the sequences discussed in
 Fig.~\ref{Fig:Multi}. As can be seen in the lower panel of
 Fig.~\ref{Fig:Multi_Mdot}, our standard mass loss recipe
 \citep[$\dot{M}_{\rm M16}$,][]{Bertolami2016} leads to a reduction of
 the mass of the star of less than 10\% during the O-rich AGB
 phase. As a consequence, a reduction in the wind prescription only
 leads to a very slight decrease in the pollution of the surface by
 dredged-up carbon, and all sequences transform into C-rich objects at
 the same time ($t\sim 967.8 $Myr, see upper panel in
 Fig.~\ref{Fig:Multi_Mdot}). Clearly, this experiment demonstrates
 that the tensions arising from the short M-star lifetimes  of
   intermediate mass stars (second panel from top of
 Fig.~\ref{Fig:Multi}) cannot be resolved by invoking lower wind
 efficiencies.  In addition, while longer C-star lifetimes can be
   reproduced by decreasing the mass loss rates, due to the immunity
   of the M-star lifetimes to mass loss this would lead to a very
   serious discrepancy in the C-star fraction, as shown in
   Fig. \ref{Fig:Multi}. Needless to say, the use of more intense
 winds would lead to a faster reduction of the H-rich envelope mass
 and to a more intense pollution by C, leading to a shortening of the
 O-rich TP-AGB phase.

 However, an alternative explanation for the previous discrepancy
  might exist, which is related to the so-called AGB boosting effect
  described by \cite{2013ApJ...777..142G}. Although in principle the
  AGB boosting effect should affect stellar lifetimes derived for
  clusters in the very narrow age range between $\sim 1.57$ Gyr and
  $\sim 1.66$ Gyr (corresponding to initial masses in the range
  between $\sim 1.75 M_\odot$ and $\sim 1.8 M_\odot$) for the LMC, the
  actual impact in the stellar lifetimes derived by
  \citep{Girardi2007} is broader due to the adopted binning
  strategy. Specifically, the clusters NGC 1651, NGC 1652, NGC 1751,
  NGC 1846, NGC 1856, NGC 1978, NGC 2154, NGC 2173, and NGC 2231 in
  the LMC (as well as clusters NGC 151, NGC 411, and NGC 419 in the
  SMC) might be affected by AGB boosting, and due to the binning
  strategy these clusters are included in both the 1.66 $M_\odot$ and
  2.17 $M_\odot$ bins for the LMC in \citep[see Table 2
    of][]{Girardi2007}. In particular the LMC lifetimes for 2.17
  $M_\odot$ adopted for the discussion of the present section could be
  dominated by clusters affected by the AGB boosting effect, strongly
  reducing the previously discussed discrepancy. Marigo \& Girardi
  (2019, private communication) estimate that lifetimes in these
  particular bins of \citep{Girardi2007} might be overestimated by up
  to a factor 2, although a detailed study is required to assess
  whether this is the case or not. 

\section{Concluding remarks}
\label{sec:remarks}
We have performed an extensive exploration of the impact of Convective
Boundary Mixing (CBM) on the predictions of stellar evolution models
for the Thermally Pulsating AGB phase (TP-AGB). In particular we have
not only explored the impact of CBM during the TP-AGB but also the
impact of CBM during previous evolutionary stages on the predictions
of TP-AGB stellar models. Our study shows that CBM on the pre-AGB
phase has a significant impact on the evolution of TP-AGB models,
affecting TDU, HBB and also the IFMR of the models. This result
indicates that  studies that aim to link AGB or PNe abundances to
  the initial stellar progenitor masses, or ages of the host
  population, should rely on stellar models that include CBM at all
pre-AGB phases. This is in agreement with
  recent hints of HBB taking place in stars with initial masses as low
  as $M_{\rm ZAMS}\sim 3  M_\odot$\citep{2018MNRAS.473..241H,2018ApJ...853...50F,2019arXiv190908007D}.

As part of this study, we have explored a recent argument by
\cite{Lattanzio2017} that buoyancy prevents momentum-driven overshoot
at the base of the PDCZ, and shown that, while this is true, this
argument also suggest that momentum-driven overshoot cannot happen at
the bottom of the CE, where it is necessary to form the $^{13}$C
pocket required for s-process nucleosynthesis. Therefore, we also
performed an estimation of the 
intensity of turbulent entrainment during thermal pulses both at the
bottom of the PDCZ and the CE and proved that, contrary to what happens
to momentum-driven overshoot, turbulent entrainment seems to be quite
capable of producing CBM on the TP-AGB.

 We confirm here that oxygen surface abundances in
  PG1159-type stars can be reproduced with stellar models that include
  CBM at the PDCZ with values \fPDD=$0.004$--$0.014$, although the
  large spread in individual PG1159 abundances also suggest the
  existence of some object-specific mixing, like rotation-induced
  mixing, might also be required. Our experiments also show that the C-star
  fraction derived by \citep{Girardi2007} is better reproduced by
  models that include CBM at the bottom of the CE, and have a lower
  intensity of C-rich winds than adopted in our standard models
  \citep{Weiss2009, Bertolami2016}.  Regarding the CBM during the
  TP-AGB we have found that the combined influence of $f_{\rm PDCZ}$
  and $f_{\rm CE}$ on the third dredge-up efficiency behaves as the
  linear superposition of the individual influences of CBM at both
  convective borders (Fig.~\ref{Fig: TP11 lambda CE pdczDUB}).

  These results mentioned above show that CBM on the TP-AGB can be
  calibrated to reproduce some key observables.  Unfortunately, a
  very different picture arises when the absolute duration of the
  TP-AGB phase is considered. Stellar evolution models that include
  CBM during the TP-AGB predict systematically short TP-AGB
  lifetimes. This is mostly due to the development of very efficient
  third dredge-up episodes that considerably shorten the lifetime of
  the TP-AGB star as an O-rich object. This is in agreement with the
  results recently published by \cite{2019MNRAS.485.5666P}.

We have explored many different observables both on the AGB and
post-AGB phases and shown that strong tensions arise that make it
impossible for the stellar models to simultaneously reproduce all of
them. More specifically, stellar evolution models that include CBM in
the way of an exponentially decaying diffusion coefficient are unable
to simultaneously reproduce the M- and C-star lifetimes and C-star
fractions observed in the clusters of the LMC, the O-abundances
measured at the surface of PG1159 stars, the size of the $^{13}$C
pocket as required from s-process nucleosynthesis and the IFMR as
determined from semi-empirical methods.  One of the key
  observables in this regard is the length of the TP-AGB phase, as
  derived from stellar counts in LMC clusters \citep{Girardi2007}. It
  is possible that TP-AGB lifetimes coming from number counts in
  globular clusters of the LMC are overestimated by up to a factor of
  two in the case of the $M_i=$ 1.66 $M_\odot$ and 2.17 $M_\odot$ bins due to
  the cluster binning and the AGB boosting effect
  \citep{2013ApJ...777..142G}. If this is the case, most of this
  tensions would be signifficantly reduced. On the contrary, if the
  relatively long TP-AGB lifetimes derived by \citep{Girardi2007} are
  supported by future studies this would cast doubt on our current
ability to produce accurate models for the AGB and post-AGB phases, as
well as to produce reliable yields from AGB nucleosynthesis.
Alternative CBM recipes need to be developed and tested for the
TP-AGB. In particular time-dependent mixing schemes based on the
convective entrainment picture of \cite{Meakin2007}, and the impact of
rotationally induced mixing on the TP-AGB are worth exploring.

\section*{Acknowledgments}
 We thank the referee for her/his very detailed reports and for
  pointing to us the possible impact of the AGB boosting effect in our
  analysis.  This work was enhanced due to the
Deutscher Akademischer Austausch Dienst (DAAD) and was supported by
the DFG-Cluster of Excellence ``Ursprung und Struktur des
Universums''.  M3B is partially supported by ANPCyT through grant
PICT-2016-0053, and by the MinCyT-DAAD bilateral cooperation program
through grant DA/16/07. M3B also thanks the visitor program from the
Max-Planck-Institut f\"{u}r Astrophysik.  This research has made use
of NASA's Astrophysics Data System.

\bibliographystyle{mnras}
\bibliography{ref} 





\bsp	
\label{lastpage}
\end{document}